\documentclass[10pt, journal]{IEEEtran}
\IEEEoverridecommandlockouts

\def\BibTeX{{\rm B\kern-.05em{\sc i\kern-.025em b}\kern-.08emT\kern-.1667em\lower.7ex\hbox{E}\kern-.125emX}}

\usepackage{nicefrac}
\usepackage{siunitx}
\usepackage{array,framed}
\usepackage{booktabs}
\usepackage{
  color,
  float,
  epsfig,
  wrapfig,
  graphics,
  graphicx,
  subcaption
}
\usepackage{textcomp,amssymb}
\usepackage{setspace}
\usepackage{latexsym,fancyhdr,url}

\usepackage{enumerate}
\usepackage{adjustbox}
\usepackage[ruled,vlined]{algorithm2e}
\usepackage{algpseudocode}
\usepackage{graphics}
\usepackage{xparse} 
\usepackage{xspace}
\usepackage{multirow}
\usepackage{csvsimple}
\usepackage{balance}
\usepackage{hyperref}
\usepackage{xurl}

\usepackage{import}
\usepackage{xcolor}
\usepackage{color, soul}

\usepackage{
  tikz,
  pgfplots,
  pgfplotstable
}
\usepackage{hyperref}

\usetikzlibrary{
  shapes.geometric,
  arrows,
  external,
  pgfplots.groupplots,
  matrix
}

\pgfplotsset{compat=1.9}

\usepackage{mathtools,}

\DeclareMathAlphabet{\mathcal}{OMS}{cmsy}{m}{n}

\DeclareGraphicsExtensions{%
    .png,.PNG,%
    .pdf,.PDF,%
    .jpg,.mps,.jpeg,.jbig2,.jb2,.JPG,.JPEG,.JBIG2,.JB2}

\usepackage{amsmath,amssymb,amsfonts}
\usepackage{graphicx}
\usepackage{textcomp}
\usepackage{xcolor}
\usepackage{algcompatible}
\usepackage{algpseudocode}
\usepackage{setspace}
\usepackage{listings}
\usepackage{caption}
\usepackage{comment}
\usepackage{enumitem}
\usepackage{multirow}
\usepackage{lipsum}

\definecolor{black}{RGB}{0,0,0}
\definecolor{function}{RGB}{0,102,153}      
\definecolor{lightgreen}{RGB}{102,153,0}    
\definecolor{bluegreen}{RGB}{51,153,126}    
\definecolor{magenta}{RGB}{217,74,122}  
\definecolor{orange}{RGB}{226,102,26}       
\definecolor{purple}{RGB}{125,71,147}       
\definecolor{green}{RGB}{113,138,98}        

\lstdefinelanguage{Verilog}{
morekeywords = [1]{assign},
keywordstyle = [1]\color{red}
}

\lstdefinestyle{interfaces}{
  float=tp,
  floatplacement=tbp,
  abovecaptionskip=-5pt
}
\usepackage[noadjust]{cite}

\usepackage{soul}

\date{}

\begin{document}
\title{Hardware-Assisted Detection of Firmware Attacks in Inverter-Based Cyberphysical Microgrids} 


\author{\IEEEauthorblockN{Abraham Peedikayil Kuruvila*,~\IEEEmembership{Student Member,~IEEE}, Ioannis Zografopoulos*,~\IEEEmembership{Graduate Student~Member,~IEEE},  Kanad Basu,~\IEEEmembership{Senior~Member,~IEEE}, and   Charalambos Konstantinou,~\IEEEmembership{Senior~Member,~IEEE}
}

\thanks {* Equal contribution, ordered alphabetically.}
}

\maketitle

\IEEEaftertitletext{\vspace{-1\baselineskip}}

\begin{abstract}
The electric grid modernization effort relies on the extensive deployment of microgrid (MG) systems. MGs integrate renewable resources and energy storage systems, allowing to generate economic and zero-carbon footprint electricity, deliver sustainable energy to communities using local energy resources, and enhance grid resilience. MGs as cyberphysical systems include interconnected devices that measure, control, and actuate energy resources and loads. For optimal operation, cyberphysical MGs regulate the onsite energy generation through support functions enabled by smart inverters. Smart inverters, being consumer electronic firmware-based devices, are susceptible to increasing security threats. If inverters are maliciously controlled, they can significantly disrupt MG operation and electricity delivery as well as impact the grid stability. In this paper, we demonstrate the impact of denial-of-service (DoS) as well as controller and setpoint modification attacks on a simulated MG system. Furthermore, we employ custom-built hardware performance counters (HPCs) as design-for-security (DfS) primitives to detect malicious firmware modifications on MG inverters. The proposed HPCs measure periodically the order of various instruction types within the MG inverter's firmware code. Our experiments illustrate that the firmware modifications are successfully identified by our custom-built HPCs utilizing various machine learning-based classifiers.
\end{abstract}

\begin{IEEEkeywords}
Microgrids, smart inverters, firmware attacks, hardware performance counters. 
\end{IEEEkeywords}

\section{Introduction} \label{sec:intro}

The electric power grid is transitioning from an utility-centric and hierarchical system to a decentralized, dynamic, and intelligent grid. This transformation relies on incorporating renewable resources, microgrids (MGs), demand-response mechanisms, smart metering infrastructure, and distributed energy resources (DERs). Renewable portfolio standards promote the adoption of DERs including energy storage systems, electric vehicles, solar photovoltaic (PV) plants, and grid-tied solar inverters. For instance, California has a goal to generate 50\% of its power demand using only renewable resources by 2030 \cite{Cal_RPS}.
According to the International Energy Agency (IEA), the globally distributed solar PV capacity is expected to increase more than 250\% in 2019-2024, reaching 530~GW \cite{iea}. Furthermore, by 2050, the forecasted solar and wind generated capacity will account for almost 30\% of the U.S. electricity demand \cite{electrify}.

The rapid growth of the solar energy market is enabled by the widespread deployment of solar inverters.  A solar or PV inverter is an electric converter which converts the variable direct current (DC) output of a PV solar panel into the utility frequency alternating current (AC) and thus supplies the grid or uses it locally off the grid.  
Contrary to traditional plants, solar PV plants and in general DERs are operated by end-users and aggregators; hence,  utilities have limited control of the power generation. For instance, residentially deployed solar inverters (e.g., in home installations) are considered consumer electronic devices~\cite{ZHOU201630}. Inverter management systems enable users to analyze energy consumption and generation patterns, in order to optimally control home energy expenditure and minimize operational cost~\cite{han2014smart}. However, to harness these features and fine-tune inverter performance, constant communication with the utility grid is required to ensure current information (e.g., energy prices, demand-response schemes, etc.).

The IEEE 1547 interconnection standard specifies the DER communication requirements as well as the control functions which should be provided to independent system operators (ISO) and distribution system operators (DSO)~\cite{IEEE1547}. 
Despite the importance of ensuring compliance with the recommendations of IEEE 1547 -- as it can significantly reduce threats targeting the communications and control commands issued between utilities and DERs -- cyberattacks that can disrupt the normal operation still remain as potential threats. For instance, adversaries can spoof DER-to-utility communications to perform denial-of-service (DoS) attacks or false data injection attacks severely impacting the distribution system operation. 
Therefore, ways to detect and mitigate such attacks have been investigated in the literature \cite{sayghe2020survey}. For instance, network segmentation could minimize potential attack entry points, however, it can still introduce issues. Enforcing firewall rules, virtual private networks (VPNs), secure enclaves, etc. help monitoring bidirectional traffic. However, such methods increase latency, network administration overheads, and in the case of DER networks (which are constantly expanding, integrating new devices and are not centrally controlled or owned by one entity), they are often infeasible~\cite{roadmap}. Only recent works have proposed ways to address the computation overheads induced by securing DER communications. For instance, the authors in~\cite{zografopoulos2020derauth} propose a lightweight hardware-based security primitive leveraging real-time grid device entropy to enable secure communication between power system assets. They validate the framework practicality in a simulation environment where DER and grid devices utilize IEEE 1815-Distributed Network Protocol v3 (DNP3) for secure authentication purposes~\cite{IEEE1815,zografopoulos2020harness}.

Apart from attacks that target the availability of DER assets through their communication functionalities, malicious attacks can also target the actual physical components of DER devices such as smart solar (micro-)inverters and their controllers. Following the same trend with most consumer electronics, manufacturers typically design DER embedded controllers using commercial-off-the-self (COTS) components, and therefore, vulnerabilities of these modules can also be exploited in such critical devices. In addition, DERs' control units often have limited computing capabilities relying on low-level hardware systems for their operation without supporting operating systems~\cite{remoteDanger}.  They typically boot monolithic, single-purpose firmware, and the tasks are executed in a single-threaded super-loop based execution. 

\begin{figure}[t!]
\centerline{\includegraphics[width=0.95\linewidth]{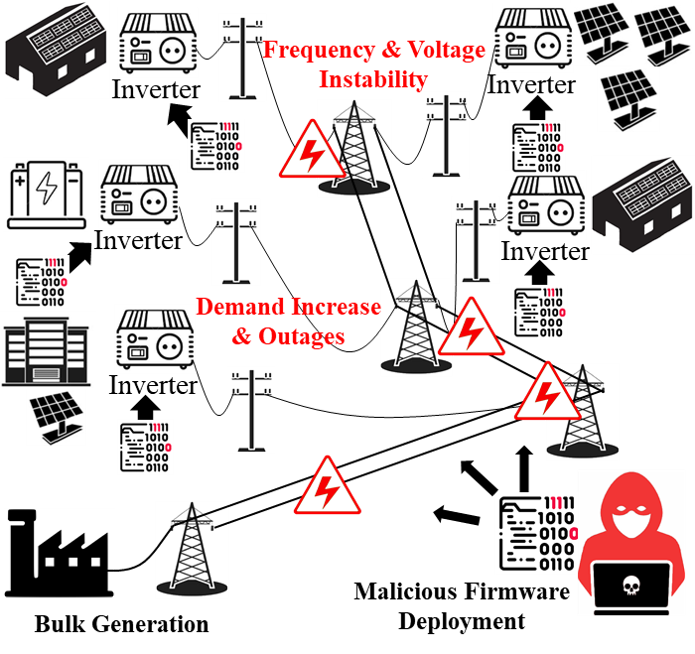}}
\caption{Malicious firmware deployment in a DER-enabled power grids.}        \label{fig:overview}
\end{figure}

Furthermore, to minimize cost, manufacturers overlook the security of these controller devices \cite{stright2020defensive}. Software patches are typically issued to address security flaws post-deployment. Such updates can be performed manually, i.e., by the device user or certified personnel, using the wired network that the device is connected to, or over-the-air (OTA) for wirelessly connected devices. In~\cite{9107609}, the authors demonstrate that attackers can intercept data in-transit during OTA update procedures, modify them, and update (remotely) malicious firmware images to programmable controller devices compromising their operation. In order to address the evident vulnerability of firmware updates from untrusted sources,  the authors of~\cite{falas2020modular} provide a streamlined approach for secure firmware updates. However, such solutions have not yet been deployed in industrial and commercial embedded systems. Contrary to industrial controllers and other embedded devices of cyberphysical systems which might be protected by sophisticated security perimeters (e.g., firewalls, VPNs, network enclaves, demilitarized zones, etc.), smart inverters are often vulnerable against adversaries with limited system knowledge and resources~\cite{qi2016cybersecurity, 8975537}. Thus, attacks targeting the firmware of such embedded systems, as outlined in Figure~\ref{fig:overview}, can cause severe impacts to the grid's critical infrastructure. 
A recent example is the Trisis incident in 2017 targeting petrochemical plants in Saudi Arabia. The attackers were able to compromise six emergency shut down controllers managing critical industrial processes (e.g., burner management)~\cite{Trisis1}. Human lives would have been in danger if the attack was not timely identified, and harmful gases were released in the petrochemical facility.

In order to mitigate the inherent vulnerabilities of embedded controllers including smart inverters, anti-virus software (AVS) has been utilized for the detection of malicious code. However, attackers have been able to evade AVS through diligently constructing attacks. For example, attackers can create viruses that modify the appearance of the code while maintaining malicious functionality~\cite{murad2010evading}.

Thus, a cat-and-mouse race between attackers and AVS ensues. Attackers attempt to develop novel methods for evading AVS, while AVS seeks to fortify its defenses. The persistent bolstering of AVS leads to large computational bandwidth and performance overhead. Furthermore, AVS cannot always identify polymorphic and metamorphic malware that can change their execution format in every iteration of propagation~\cite{demme2013feasibility}. To address the issues arising from AVS, researchers have proposed hardware malware detectors (HMDs)~\cite{malone2011hardware, ozsoy2016hardware}.

HMDs utilize hardware features which, compared to AVS, are more robust since compromising them is extremely challenging~\cite{demme2013feasibility}. Moreover, since HMDs operate on actual hardware, the malware detection latency is lower.
Hardware performance counters (HPCs) have emerged as a promising candidate for HMDs. HPCs are dedicated registers that keep track of low-level microarchitectural events such as the number of branch-misses, CPU cycles, instructions, etc. Towards malware detection, HPC values,  collected from the execution of applications, can be used to train machine learning (ML) models, which can classify the programs as benign or malicious~\cite{sayadi20192smart, wang2016malicious}. Despite the detection and performance benefits of HPC-based HMDs,
directly applying it to MG systems is not feasible, since some embedded legacy controllers do not support HPC.      

In this paper, we leverage custom-built HPCs as design-for-security (DfS) primitives for firmware-based solar inverter controllers. These custom-built HPCs keep track of the instruction sequences (inside the controller firmware). We utilize ML classifiers to differentiate between maliciously modified and benign firmware versions. Specifically, our contributions are as follows: 

\begin{itemize}
    \item We design firmware modification attacks for solar inverters considering their characteristics and how they operate in a MG setup,
     \item We assess the impact of these attacks targeting smart inverter controllers on a simulated MG architecture, 
     \item We design custom-built HPCs to enable the detection of malicious firmware modifications for such inverter controllers that do not support HPC functionality, and,
    \item We utilize different ML classifiers to detect the firmware modification attacks using our custom-built HPCs values.
\end{itemize}


The rest of the paper is organized as follows. Section~\ref{s:background} presents the background and some preliminary definitions. Section~\ref{RelatedWork} 
describes related work on power systems cybersecurity and HPC-based malware detection. Section~\ref{s:Methodology} describes our proposed methodology, and  Section~\ref{sec:simulation} presents the experimental results. Finally, Section~\ref{sec:Conclusion} concludes the paper and gives directions for our future work.

\section{Background}\label{s:background}

The following subsections include short descriptions of power systems preliminaries as well as background on HPCs and  ML classifiers used in this study.
\subsection{Power System Preliminaries}

The next generation of the power grid, often referred to as smart grid, leverages information and communication technologies in order to efficiently and reliably supply electricity~\cite{NISTsmartgrid, smartgrid}. Contrary to traditional power systems where electricity flows in one-way from bulk generation facilities to consumers and loads, smart grid enables a bidirectional flow of energy and information between prosumers, consumers, and utilities. A MG is a group of generation resources (e.g., natural gas or biogas generators,  wind turbines, solar PV, etc.), energy storage systems (ESS), and loads (residential, commercial, or/and industrial), which can operate either in connected or autonomous/islanding mode. MGs normally operate connected to and synchronous with the grid, but they can also function autonomously if necessary (e.g., extreme weather events, utility load shedding to maintain stability, etc.). MGs retain all the smart grid characteristics, including bidirectional energy flow and communication. MGs also serve as buffers between the MG-integrated devices and the actual grid. Utilities, instead of independently communicating and controlling each device on the distribution level (e.g., distributed generators, inverters, ESS, loads, etc.), hand over control to MG management systems. MGs help reduce significant communication delays, costs, and computation resource overuse, by serving as a nexus between utilities and devices~\cite{arbab2019smart}. We will leverage a similar approach in our impact analysis; the specific details of our MG setup are presented in Section~\ref{s:MGmodel}.

Different inverters 
exist in order to accommodate the field application requirements of grid systems~\cite{Inv1, Inv2}. For instance, depending on the generation source, inverters can be categorized as solar, battery, thermoelectric, and hybrid inverters. Solar inverters are electronic devices which enable the conversion of DC power -- generated by the PV panels -- to AC power which can be fed back to the power grid or can be supplied to local appliances. Additionally, based on their output characteristics, inverter technologies can be of square wave, sine wave, or modified sine wave~\cite{Inv3}. Depending on the connection practice, inverters are also classified as off-grid or grid-tied. In this work, we focus on \emph{solar, pure AC sine, and grid-tied microinverters} which are the most prevalent topology in MGs~\cite{InvStatus}. 

The main difference between standard solar inverters and microinverters is that for the former, the output of an array of interconnected solar panels is provided to a centralized solar inverter, while in the microinverter case, every panel is connected to its own microinverter. The main disadvantage of standard inverters is that the generated power is indicated by the least efficient solar panel. Microinverters are optimally designed to take full advantage of the connected PV panel reaching power conversion efficiency of up to 96\%~\cite{InvEf}. This is achieved using the maximum power point tracking (MPPT) mechanism embedded in the microinverter, which ensures that the maximum available power can be extracted from the PV panel under the varying environmental conditions (e.g., shading, high ambient temperature, etc.). In the rest of the paper, the terms inverter and microinverter are used interchangeably, without loss of generality, since similar principles dictate their operation.

The dependency of microinverters on COTS components provides the ``smart'' functionalities of modern embedded devices such as remote OTA updates, communication and control support functions, and high-speed data sampling and acquisition.  However, this also incurs security vulnerabilities~\cite{smartDanger, liu2020deep}. In addition to being deployed in insecure or unsupervised environments, these controllers were not initially developed with security in mind, and their resource-constrained architectures provide limited options for improvements in this direction. Multiple attacks targeting controllers in industrial control systems and critical infrastructure have been reported in literature  highlighting the need for potent attack detection frameworks~\cite{ICS_attack, mclaughlin2016cybersecurity}.

\subsection{Hardware Performance Counters}
\label{sec:HPCs}
In order to detect firmware attacks on MG systems, we utilize HPCs. HPCs are special-purpose registers, found in most modern processors that monitor critical low-level microarchitectural events like the number of cache-misses, branch-misses, instructions, etc., developed to improve system performance. Typically, HPC collection is attained through operating system packages such as {\it perf} in Linux systems, or other software binaries like {\it quickhpc } utilizing the {\it PAPI } framework. The number of HPCs that are accessible as well as the number of HPCs that can be monitored simultaneously differ from processor to processor. Some processors only support a couple of HPCs, while others support tracking of multiple events. In situations where HPCs are unavailable or limited, custom HPCs can be designed to keep track of features that encapsulate a program's functionality. In the recent past, HPCs have been utilized along with ML classifiers to detect malicious applications~\cite{demme2013feasibility, wang2016malicious}.

\subsection{Machine Learning Classifiers}
In this part, we furnish a brief explanation for the ML classifiers utilized in our experiments. We also present a background on the principal component analysis (PCA) used to reduce the number of features. Finally, we explain the various metrics used for quantifying the performance of the utilized ML algorithms. 

\subsubsection{ML Classifiers}
\label{MLclassifiers}

A {\it decision tree}~(DT) is an algorithm where the classification model that is built is in a top-down recursive tree like structure, using mutually exclusive if-then rule set. The rules for the model are sequentially learned one at a time, based on the initial training dataset~\cite{asiri_2018}. 
A {\it neural network} (NN) is a ML classifier that utilizes a set of neurons and layers. Neurons are computational units  interconnected between each other, and every connection contains a weight~\cite{HowtoCon88:online}. The classifier learns and adjusts weights as the model is trained to correctly classify the proper label for the input tuples. 
A {\it random forest} (RF) is a model that operates as an ensemble incorporating many DTs. For classification problems, the model functions by having each DT furnish a prediction, and the output of the RF classifier is the prediction that is most selected~\cite{Understa84:online}. This model operates on the concept that a group of uncorrelated trees will provide better results as opposed to utilizing a single tree.

\subsubsection{PCA for Feature Selection}
\label{PCA-explanation}

Although processors can include multiple HPCs, only few of them can be monitored simultaneously. In order to reduce the set of HPCs (being monitored), we use PCA, a feature selection technique managing high dimensionality data with multiple variables. 
PCA is a data reduction technique allowing the user to specify the number of principal components in the transformed data.  PCA first calculates the covariance matrix which is formed from the initial dataset of features. The covariance matrix contains all possible combinations of variance between two variables. Singular value decomposition is then used to factorize the matrix and extract data in the directions with the highest variances~\cite{MachineL58:online}. Based on the specified principal components, the values in the matrix correspond to the features in the initial dataset. The best features aree selected based on the largest values. 

\subsubsection{ML Terminology and Metrics}

In this section, we furnish the definitions for some ML terminology and measurement metrics utilized in this work. A true positive (TP) is a malicious application correctly labeled malicious and a true negative (TN) is a benign application accurately classified as benign. A false positive (FP) is a benign application incorrectly classified as malware, while a false negative (FN) is a malware mistakenly tagged benign.

\textbf{Accuracy} is the proportion of the number of predictions correctly classified to the total number of predictions made. \\
\begin{equation}
\label{equ1}
Accuracy =  \frac{No.\  of\ (TP\ +\ TN)}{ No.\  of\ (F  P\ +\ F N\ +\ T P\ +\ T N)\ }
\end{equation}

\textbf{Precision} represents the ratio of positive class classifications that are correct. \\
\begin{equation}
\label{equ2}
Precision =  \frac{No.\  of\  T  P}{No.\ of\ F  P\ + No.\ of\ T P\ }
\end{equation}

\textbf{Recall} is the proportion of correct positive classifications to the total number of positive classifications.\\
\begin{equation}
\label{equ3}
Recall =  \frac{No.\  of\  T  P}{No.\ of\ F  N\ + No.\ of\ T P\ }
\end{equation}

\section{Related Work} \label{RelatedWork}

The rapid integration of DERs and smart inverters in the power grid, besides the operational benefits, raises many security concerns. Cyberattacks targeting embedded systems and programmable controllers have been launched with varying degrees of impact, ranging from economic losses and power outages to even life-threatening scenarios~\cite{Attack1, stuxnet}. 
Research has emphasized the impact that cyberattacks targeting smart grid assets, e.g., smart inverters, can have on MGs  as well as  the power system in general~\cite{zografopoulos2021cyberphysical, DER2}. Despite the cybersecurity standards and mitigation strategies which have been proposed to enhance grid stability, implementing such approaches results in performance overheads and often requires specialized equipment which cannot be retrofitted to deployed legacy devices or ported to proprietary architectures~\cite{DER0, DER1}. { For instance, security solutions leveraging trusted platform modules (TPMs), secure boot, and cryptographic signatures used to validate the integrity of firmware updates, could address some of the aforementioned issues. However, they incur an additional overhead of extra hardware modules and extensive redesign of the deployed and resource-constrained systems. Thus, novel vendor-agnostic methods -- utilizing the inherent embedded system infrastructure -- are essential to provide better situational awareness and aid the early detection and recovery of abnormal system operation}.

Prior works have shown the capabilities of defending against malicious firmware. The authors in~\cite{li2011viper} proposed a software-only attestation method, VIPER, that checks the peripherals' firmware integrity. This defense approach assumes that the host CPU and operating system have not been compromised. The work in~\cite{maskiewicz2014mouse} utilized a signature verification code method in the bootloader, which is assumed to have sufficient memory space, to prevent malicious firmware from being installed onto a mouse. In~\cite{lemay2009cumulative}, the authors proposed a cumulative attestation kernel architecture consisting of flash microcontroller units for remote attestation assurance that allowed the inspection of the integrity of an application's firmware. Other works like~\cite{abad2013chip} utilized the addition of an on-chip control flow monitoring module that detects malicious firmware by checking the control flow integrity in a system. However, this technique incurs high performance and storage overhead. 

In this work, we use HPC-based AVS for detecting malicious firmware modifications of MG inverters. Utilization of HPCs for detecting malware was initially proposed by~\cite{malone2011hardware, wang2015reusing}, which was improved upon by applying ML models in conjunction with HPCs for application classification~\cite{demme2013feasibility}. 
{The work in~\cite{krishnamurthy2019anomaly} detected malware in cyberphysical systems through real-time quantification of HPCs. The study in \cite{sayadi20192smart} proposed {\it 2SMaRT}, an approach that selects the best HPCs through feature selection and applies a two-stage classifier for malware detection. In addition, the authors in~\cite{bahador2014hpcmalhunter} developed {\it HPCMalHunter} which produces behavioral vectors for programs created from the HPCs for real-time malware detection. } Furthermore, the works in~\cite{wang2015confirm, wang2016malicious} extended utilizing HPCs in embedded control systems to detect malicious firmware modifications through a comparison-based approach. Since some systems lack adequate HPC capabilities, various methods of custom HPCs have been proposed in prior works for malware detection. Researchers have defined and explored detecting malicious applications through ML models trained on sub-semantic features~\cite{ozsoy2015malware, ozsoy2016hardware}. On the other hand, custom HPCs that count the ordering of specific instructions at the assembly level are used to train ML classifiers capable of detecting morphing malware~\cite{rohan2019can}.                          
\section{Proposed Methodology}\label{s:Methodology}

In this section, we provide an overview of our threat model as well as the critical components of the MG inverter controller and the firmware attacks we designed to disrupt MG operations. Furthermore, this section describes our proposed custom HPC-based methodology to detect firmware modifications. 

\vspace{-1mm}
\subsection{Threat Model}
Our threat model considers an attacker who intends to compromise the microinverter controller aiming to impact the power conversion process. In order to achieve this goal, we assume that the attacker has compiled a counterfeit firmware version and attempts to port it to the actual device \cite{7741452, konstantinou2019hardware}. Thus, the attack focal point is on the application level and aspires to trigger abnormal operations via static firmware modifications \cite{konstantinou2015impact}. The adversarial access can either be physical, by intrusively uploading the malicious firmware and affecting the microinverter operation, or through remote exploits (e.g., OTA firmware update functionality). The attacker exploits the power conversion process, thereby adversely affecting the MG operation by injecting malicious code which either manipulates the operation setpoints or alters the sensor measurements received by the controller unit \cite{10.1145/2994487.2994491}. These stealthy process-aware attacks~\cite{7523254}, targeting embedded devices and programmable logic controllers, are difficult to detect and can have various degrees of impact from equipment failures to catastrophic collapse of the industrial cyberphysical system. We demonstrate and discuss the impact of the attacks on the system in Section~\ref{sec:simulation}. {Further, we exclusively investigate custom-built security primitives which employ the innate functionalities of embedded devices without requiring sophisticated software~(e.g., cryptographic signatures), hardware~(e.g., TPMs), or compound~(e.g., secure boot) security extensions.} 


\vspace{-2mm}
\subsection{Microinverter Controller} \label{s:MPPTcontroller}

\begin{figure}[t]
\centerline{\includegraphics[width=\linewidth]{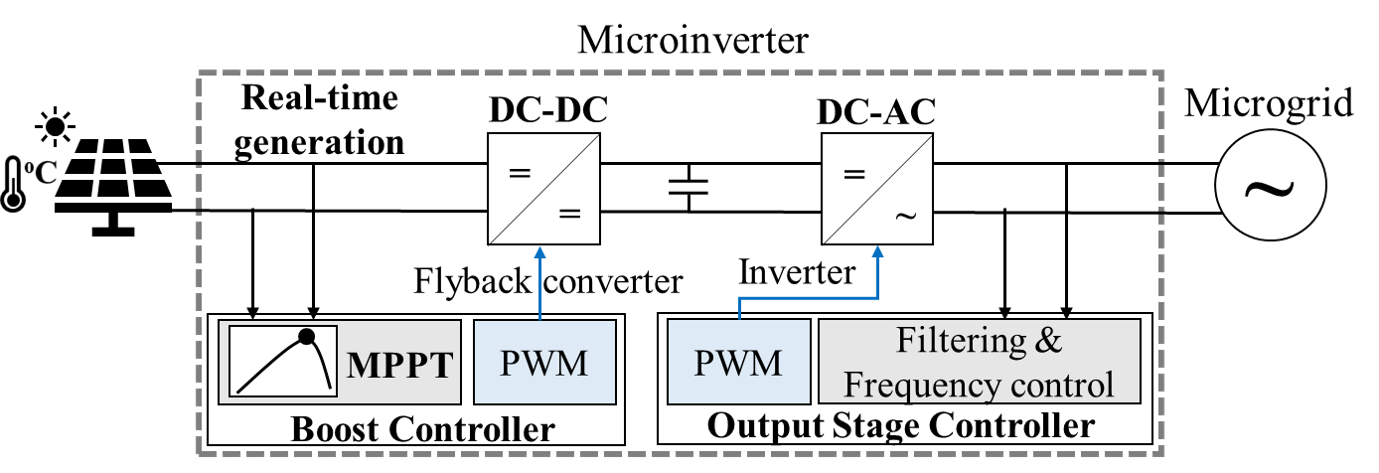}}
\caption{Microinverter architecture.}
\label{fig:methodology}
\end{figure}

The objective of a grid-tied inverter is to convert DC power, provided either from PV panels or energy storage systems, into AC power that can be sourced to the AC grid or MGs. This process is supported by CMOS devices used as switches and filters as well as voltage and frequency control modules (e.g., pulse width modulation -- PWM). The architectural components of a solar inverter are outlined in Figure~\ref{fig:methodology}. 

Microinverters are typically preferred over typical inverter topologies for residential and PV deployments in the distribution grid level ($<$ 33 $kV$) due to their higher efficiency~\cite{microPerformance}. A critical component that enables microinverters to attain high power conversion efficiency is their MPPT controller. MPPT algorithms aim to find the optimal 
voltage or current point, which will allow for the maximal power generation under the constantly changing exogenous conditions affecting the solar PV panels (solar irradiance, shading, temperature, etc.). A variety of MPPT tracking algorithms have been proposed by researchers~\cite{MPPTs}. 
Two of the most commonly utilized MPPT methodologies, which rely on a hill-climbing process to reach the optimal conversion point, are the perturb-and-observe (PnO) method and the incremental conductance scheme~\cite{PO_1}. 
Both of the 
techniques require that the PV real-time generation attributes, i.e., voltage and current, are measured periodically. In this paper, we adopt the PnO MPPT strategy. However, given the similarities between PnO and incremental conductance, our analysis can be generalized for both methods. The PnO MPPT tracker uses the sensed PV attributes (voltage $V_{rt}$, current $I_{rt}$) to generate the reference current which regulates the DC/DC flyback power stage of the microinverter~\cite{BIANCONI2013346}. An overview of the PnO operation is presented in Algorithm~\ref{alg:PnO}.

\begin{algorithm}[t!]
\setstretch{1}
\normalsize
\SetAlgoLined
\DontPrintSemicolon
Initialize $P_{i}, V_{i}$, Measure $V_{rt}, I_{rt}$ \;
 \While{MPPT.on}{
  $P_{rt}$ = Power.calculate($V_{rt}, I_{rt}$) \;
  $\Delta P = P_{rt} - P_{i}$\;
  $\Delta V = V_{rt} - V_{i}$\;
  \lIf{$\Delta P < 0$}{ 
  \\  \quad \lIf{$ \Delta V > 0 $}{ converter.currentIncrease()
      }
       \quad \lElse{ converter.currentDecrease()
      }
   } 
   $P_{i} = P_{rt}, V_{i} = V_{rt}$ \; 
 } 
\caption{Perturb and observe (PnO) algorithm.}
\label{alg:PnO}
\end{algorithm}

During operating conditions, the PV generation varies due to environmental factors, e.g., solar irradiance and temperature. The voltage-to-current relationship for each PV module differs according to its operating point, hence MPPT algorithms account for these variations. 
The power gradient is utilized in order to reach the MPPT (hill-climbing methods). By leveraging the real-time $V_{rt}$ and $I_{rt}$ measurements of the PV and the  $ \partial P$\textbackslash $\partial V$ gradient, we can determine if the MPPT point is reached, and if not, which are the necessary steps in order to approach it. Algorithm~\ref{alg:PnO} provides a discretized version of the described procedure, similar to the ones prescribed in commercial solar inverters such as the microinverter module used in our experiments.  


Since the microinverter's MPPT 
leverages the real-time PV measurements in order to extract the maximum available power from any given PV module, attackers can maliciously control the microinverter operation by stealthily tampering the sensed measurements via firmware modification attacks. Furthermore, attacks on the MPPT module can lead to large-scale impact as 
the MPPT function also controls the DC/DC boost module, which is interfaced to the DC/AC power conversion block (Figure~\ref{fig:methodology}). As a result, frequency and voltage fluctuations can be ported to the grid-tied end of the microinverter. 
In Section~\ref{sec:simulation}, we demonstrate three different attack scenarios on the MG system operation with varying degrees of impact.

\vspace{-2mm}

\subsection{Firmware Attack Designs}
\label{Firmwareattackdesigns}
Firmware attacks commonly seek to either shutdown the operation of the firmware-controlled device or cause malicious behavior that would result in an erroneous system output~\cite{konstantinou2015impact}. In this section, we discuss the set of developed attacks which emulate common scenarios that adversaries might exploit in solar microinverter code.

\subsubsection{DoS Attacks}
\label{Firmwareattackdesigns-DOS}
DoS attacks cause the system to be unavailable and not function properly. In other words, the system is temporarily or indefinitely locked, such that the typical services of a system are non-functional. In this paper, we have designed a DoS attack that switches between locking and unlocking the inverter every $10$ seconds. This attack is materialized leveraging the system timer,  and by sending interrupts 
restricting the availability of specific circuit components (e.g., microinverter output stage). In addition, we applied a diminutive version of the full system DoS attack limited to just the MPPT functional component. The ensuing ramifications on the system are shown in Section~\ref{DOSonMPPT} and Section~\ref{DOSonMI}.

\subsubsection{MPPT Input Attacks}
\label{Firmwareattackdesigns-input}
The MPPT algorithm output in the solar microinverter code controls the PV panel output current for maximum power transfer. The manipulation of the inputs to the MPPT function can cause a lack of efficiency in the grid without allowing the solar PV to operate in its full capacity. In this scenario, we focus on developing two variations of this attack. In the first attack case study, we utilize the system timer and interrupts to constantly switch between correct MPPT inputs and an array of erroneous values. These numbers are arbitrary inputs constantly switched every few interrupts. The second attack redirects MPPT inputs between correct values and the output values of a continuous sinusoidal wave every few interrupts. The resulting effects on the system are demonstrated in Section~\ref{AttackonMPPTInput}. 

\vspace{-2mm}

\subsection{HPC-Aware Firmware Modification Detection}
\label{sec:backgroundHPC}

In this part, we describe our DfS technique and propose custom-built HPCs in order to improve the security of MG inverters. In this paper, the MG inverter utilized is a solar microinverter by Texas Instruments (TI). The inverter consists of a {\it DIMM100 based controlCARD}, which is part of TI's {\it F2803x} series, that controls the inverter development board. The available profiling capabilities are extremely limited for the {\it F28x} architecture. Currently, the only profiling that can be accomplished is the count of clock cycles. This is insufficient to detect malicious firmware modifications since it does not provide any information on the structure of the program being executed. Two different applications, one benign and one malicious, being executed for the same number of clock cycles, will incur no difference in HPC values. This motivates us to design custom-built HPCs as DfS primitives of the MG system. The proposed custom-built HPCs, if integrated to next generation inverter controllers, will aid in improving the security and resilience of MG against cyberattacks. 
It has been proven that HPCs counting instructions such as load, store, arithmetic, and branch values are beneficial in capturing the dynamics of an application~\cite{rohan2019can}. Furthermore, the order of these instructions differs depending on the program being executed. With minimal resources to operate with, we have decided to investigate the development of custom-built HPCs for firmware attack detection.

\subsubsection{Custom-built HPC Design and Collection}
\label{sec:backgroundcustomHPCcollection}

The custom HPCs we propose keep track of the order of specific assembly instructions contained inside the binary executable. These HPCs are designed to include various types of instructions: arithmetic ($a$), boolean ($n$), store ($s$), load ($l$), and branch/jump ($b$). Our custom-built HPCs not only count the occurrence of each of these instruction types, but also the number of sequences of these instruction types. 

\begin{table}[t!]
\centering
\caption{Proposed custom-built HPCs.}
\vspace{0.1in}
\begin{tabular}{|c|c|c|c|c|c|c|c|c|c|}
\hline
\multicolumn{10}{|c|}{\textbf{Feature Vector}}  \\ \hline
a  & b  & l  & n  & s  & aa & ab & al & an & as \\ \hline
bb & ba & bl & bn & bs & ll & la & lb & ln & ls \\ \hline
nn & na & nb & nl & ns & ss & sa & sb & sl & sn \\ \hline
\end{tabular}
\label{featuretable}
\end{table}

\textcolor{black}{
Our HPC-based feature vector, illustrated in Table~\ref{featuretable}}, consists of 30 HPCs: $a$, $s$, $l$, $b$, $n$, $aa$, $as$, $al$, $ab$, $an$, $sa$, $ss$, $sl$, $sb$, $sn$, $la$, $ls$, $ll$, $lb$, $ln$, $ba$, $bs$, $bl$, $bb$, $bn$, $na$, $ns$, $nl$, $nb$, and $nn$. In this representation, $a$, $s$, $l$, $n$, and $b$ indicate arithmetic, store, load, boolean, and branch instructions, respectively. Therefore, the first five HPCs count the number of times each of these instructions are encountered individually. The remaining HPCs in the feature vector, in the form of $XY$, count the number  of $X$ instructions that are immediately succeeded by $Y$ instructions. As an example, HPC $la$ counts the number of load instructions that are immediately followed by an arithmetic instruction. On the other hand, HPC $al$ counts the number of arithmetic instructions that are immediately followed by a load instruction. We also propose HPCs like $bb$, which count the number of times a branch instruction follows another. As an example, Listing~\ref{lst:1} shows a sample firmware assembly code for the solar microinverter. In this code segment, HPCs $la$, $an$, $na$,  and $ab$ would all be two, while HPC $bl$ would be one. Meanwhile, HPCs $n$, $a$, $b$, and $l$ would be two, four, two, and two, respectively.

\begin{figure}[t]
\vspace{-3mm}
\centering
\begin{minipage}[t]{0.95\linewidth}
\begin{lstlisting}[caption={Snippet of firmware assembly code for solar microinverter.},label={lst:1}]
03f6438 83a1 MOV  AL,@VarA ;Load AL with  VarA
03f6439 dc18 ADD  AL,@VarB ;Add to AL VarB
03f643a da18 ANDB AL,#0xFF ;AND AL with 0xFF
03f643b dd17 SUBB XAR4, #1 ;Subtract 1 from XAR4
03f643c d918 B    404, NEQ ;Branch Not Equal
03f6438 83a1 MOV  AL,@VarB ;Load AL with  VarB
03f6439 dc18 ADD  AL,@VarA ;Add to AL VarA
03f643a da18 ANDB AL,#0xDA ;AND AL with 0xDA
03f643b dd17 SUBB XAR2, #1 ;Subtract 1 from XAR2
03f643c d918 B    253, NEQ ;Branch Not Equal
\end{lstlisting}
\end{minipage}
\vspace{-3mm}
\end{figure}

\begin{figure}[!t]
\centering
  \centerline{\includegraphics[width=70mm,scale = 0.4]{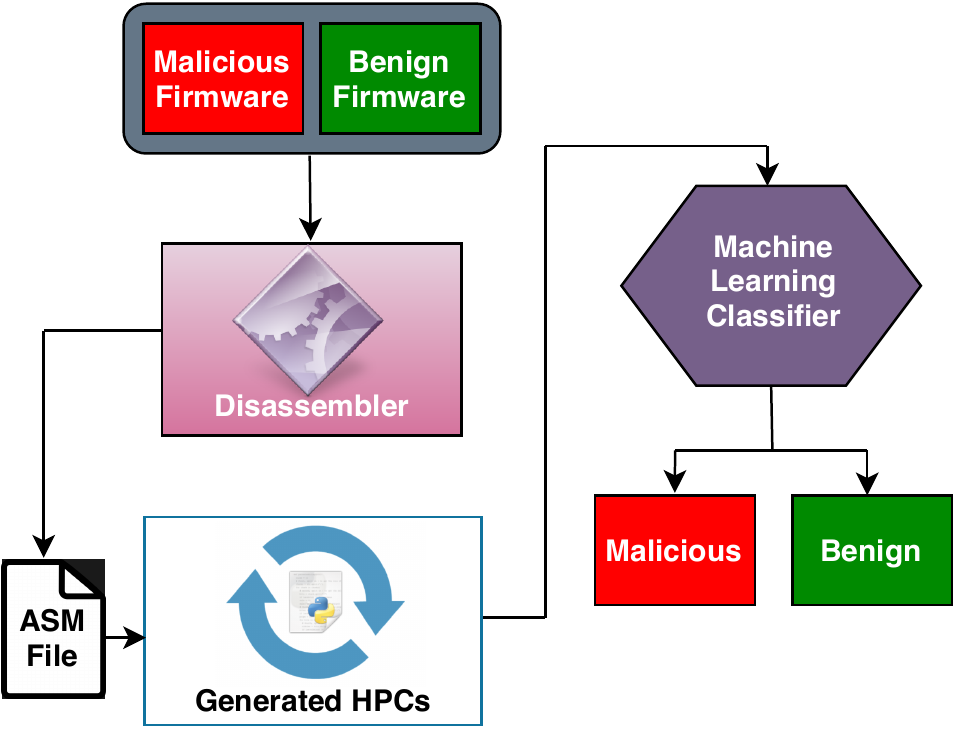}}
  \caption{Proposed HPC collection and utilization flow diagram.}
  \label{fig:HPCDiagram}
\end{figure}

The imitation of traditional HPC collection during simulation can be addressed by sampling the designed HPCs in the disassembled firmware assembly code at specific intervals. In this paper, we collect HPC counts every $50$ assembly instructions, i.e., our sampling interval is $50$. Other collection intervals can also be employed, albeit with a different number of HPC counts. The process flow graph in Figure~\ref{fig:HPCDiagram} outlines our proposed technique. First, the high-level firmware code is converted into assembly code through a disassembler. Next, we sample instructions from the assembly code periodically until we encounter $50$ instructions, i.e., our sampling interval. The values of the HPCs are recorded and the counters are reset to zero. This process is repeated until we exhaust all available assembly instructions. The recorded custom-built HPCs form a dataset, which can be used to train ML models and detect malicious firmware modifications, thereby obtaining a DfS solution against these attacks. 

Unlike traditional AVS, our DfS technique is robust against attackers that aim to obfuscate a malicious firmware through the insertion of statements at the code level. {Any insertion of additional statements will cause the disassembled firmware code to differ. Therefore, in presence of undesired instructions, the sampling of our custom-built HPCs from this assembly code will result in dissimilar values from the golden HPC trace. As a result, the ML model will be able to classify these camouflaged malicious firmware.} {The static disassembly code utilized is sufficient for distinguishing between benign and malicious firmware. Furthermore, TI does not currently provide tools for attaining dynamic traces for applications on our experimental platform.}
The proposed DfS technique is applied on a real-world commercial firmware, specifically, TI's solar microinverter that is controlled by a {\it Piccolo TMS320F28035 Isolated ControlCARD}. This DfS method can be utilized in other embedded systems where procurement of the assembly code is possible; albeit, with a difference in custom-built HPCs that can be sampled based on the available assembly instructions. Furthermore, our DfS technique is agnostic of the MG model and device. Our method does not utilize any physical measurements directly from the grid, which might be maliciously falsified~\cite{anubi2019enhanced}. The custom-built HPCs are sufficient for securing the solar microinverter because any deviations in performance or operational ranges would be engendered from malicious firmware that would be detected by our proposed defense. In addition, compared to previous works that use custom-built HPCs~\cite{rohan2019can}, our proposed technique utilizes a wider variety of features, e.g., we also incorporate boolean instructions. Furthermore, prior works have been focusing on protecting general-purpose computers, while our work is primarily concerned with securing commercial real-time power grid devices, i.e., solar microinverters.

\section{Impact Analysis and Detection Results} \label{sec:simulation}
In this section, we demonstrate the effects of firmware attacks on a MG system. We further demonstrate the efficiency of our proposed custom-built HPC-based methodology in detecting 
modifications in the microinverter's firmware code.

\subsection{Microgrid System Model} \label{s:MGmodel}

\begin{figure}[t]
\centerline{\includegraphics[width=0.9\linewidth]{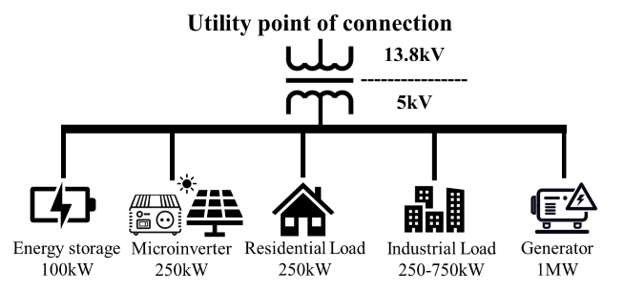}}
\caption{Structure and specifications of the simulated MG model.}
\label{fig:microgrid}
\end{figure}

In order to evaluate the impact of the firmware modification attacks discussed in Section \ref{Firmwareattackdesigns}, we develop a simulation model for the MG system. 
We utilize \emph{Matlab} 
to model the MG components including loads, generators, ESS, etc. Furthermore, we use a discrete time model for the microinverter whose MPPT controller leverages the aforementioned PnO methodology (Algorithm \ref{alg:PnO}). The grid architecture used for the MG simulation is depicted in Figure~\ref{fig:microgrid}. The simulation model of the MG includes all the components found in a realistic MG setup: a solar PV with its solar microinverter, a lithium-ion battery ESS, a diesel generator, and residential and industrial loads. The MG components are connected via a distribution substation transformer with a capacity of $250$~MVA to the utility point of connection operating at $13.8$~kV. The distribution voltage feeder level is set to $5$~kV and step-down transformers are used to interface the generation assets and the loads. The nameplate generation capacity for the diesel generator is set to $1$~MW. The maximum generation capacity for the microinverter can reach $250$~kW following the PV insolation profile. As for the ESS, it can generate up to  $100$~kW (capacity of $100$~kWh). The loads of the MG include an aggregated residential load with a constant power demand of $250$~kW and a variable lumped industrial load whose power demand ranges between $250$-$750$~kW.

\vspace{-2mm}
\subsection{Attack Impact Analysis} \label{MGImpact}

\begin{figure}[t!]
\centerline{ \includegraphics[width=0.9\linewidth]{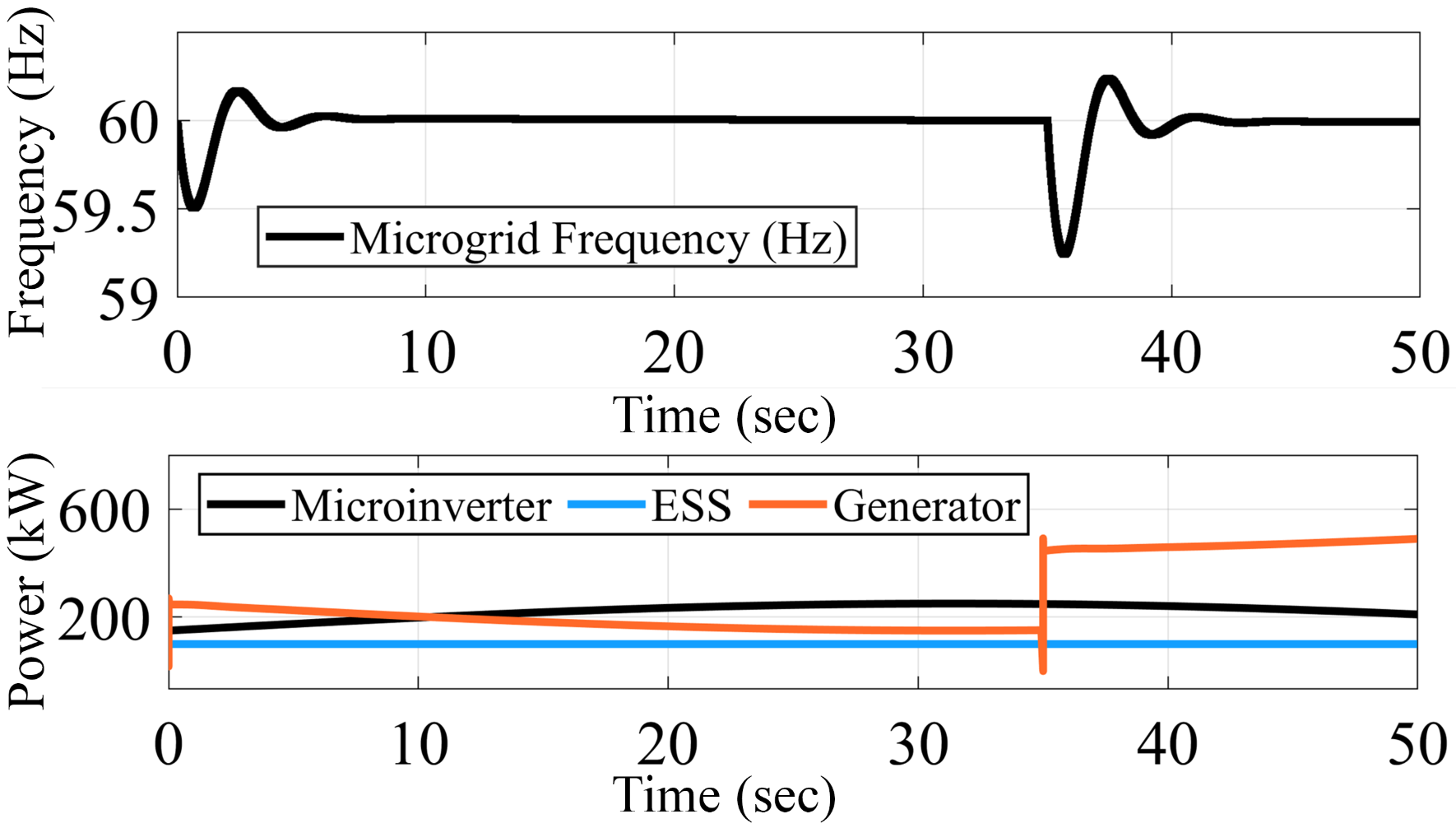}}
\caption{Nominal base case of MG behavior.}
\label{fig:nominal}
\end{figure}

The base case for our MG system is in islanded mode of operation. The islanding (or autonomous) feature of MGs is a fail-safe mechanism employed to protect both the distribution and transmission electric grid system if either one operates abnormally. In such scenario, the MG is responsible to meet the power demand of all the contained loads (e.g., industrial, residential, etc.). The islanding operation of the MG is presented in Figure~\ref{fig:nominal}. Specifically, at $t=0~sec$, the islanding command is issued, either by the DSO or MG operator, which explains the frequency fluctuation. At $t=35~sec$, we have a power demand increase (from $500$~kW to $800$~kW) in the MG -- also resulting in frequency fluctuation -- as can happen in realistic grid applications. We investigate how the MG operation is impacted under three different attack scenarios with regard to the MG operation depicted in Figure~\ref{fig:nominal}. The three attack scenarios involve: (i) a DoS attack on the MPPT controller, (ii) a DoS attack on the microinverter, and  (iii) an input-tampering attack of the MPPT controller affecting the power conversion process.

\subsubsection{DoS Attack on the MPPT Controller}
\label{DOSonMPPT}
\begin{figure}[t!]
\centerline{ \includegraphics[width=0.9\linewidth]{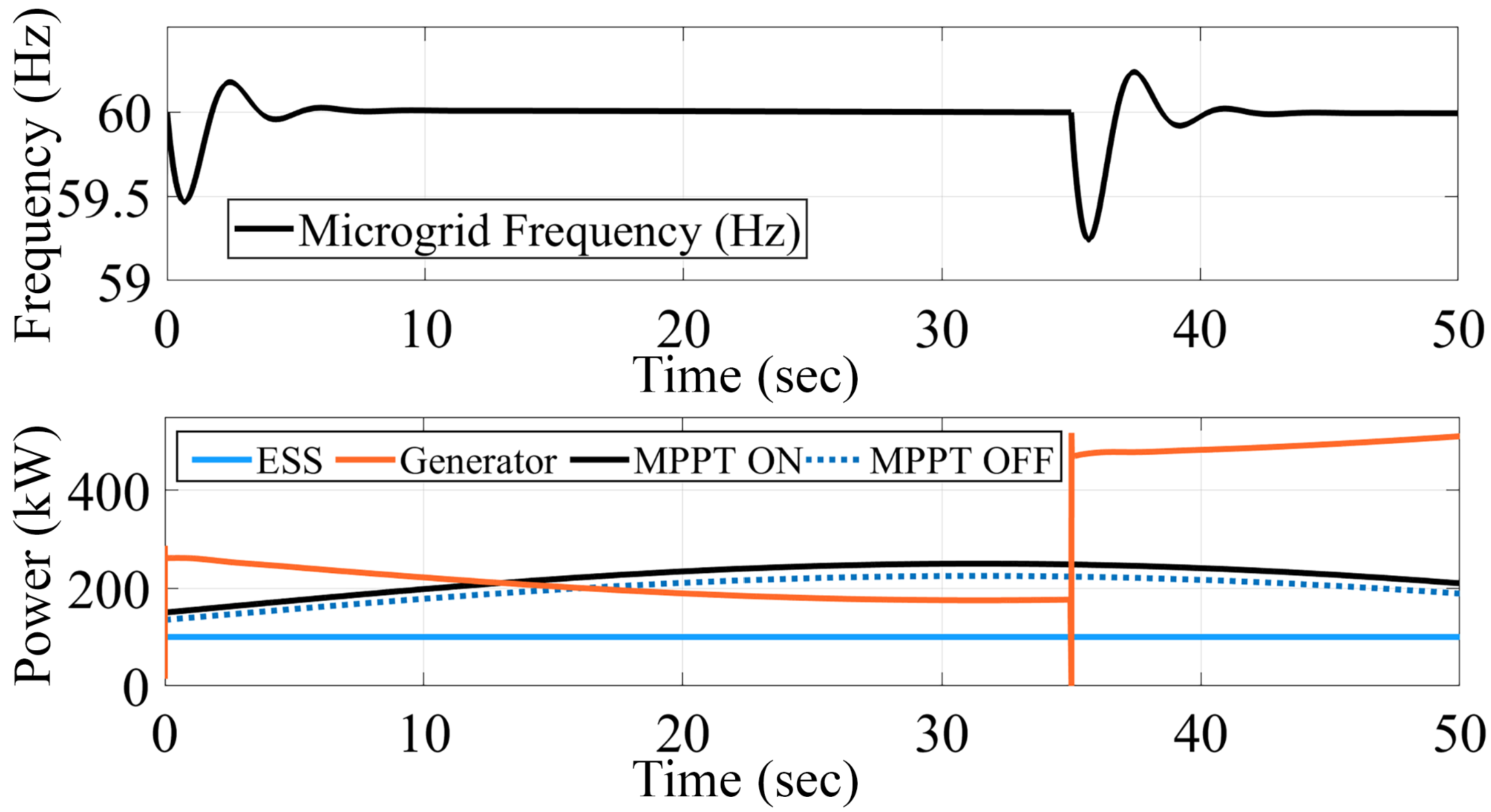}}
\caption{MG behavior under MPPT DoS attack.} 
\label{fig:MPPT_attack}
\end{figure}

In this type of attack, we modify the microinverter firmware and switch off the MPPT functionality. As explained in Section~\ref{s:Methodology},  MPPT is a feature that has been recently included in inverter architectures as it can boost their power conversion efficiency. By disabling this feature via a DoS-type of attack, malicious adversaries can expect that the generated power enabled by the microinverter will be lower. This is validated by simulation results presented in Figure~\ref{fig:MPPT_attack}. 
If the MPPT functionality of multiple inverters is disabled, thus leading to severely curtailed solar generation, it can result in brownout or load shedding events since the power reserves (e.g., alternative generation sources) might not be able to account for the power deficit. Although this type of attack might not have catastrophic impacts on the MG operation, it can elicit uneconomical power grid operation \cite{chen2013economic}.

\begin{figure}[t!]
\centerline{ \includegraphics[width=0.9\linewidth]{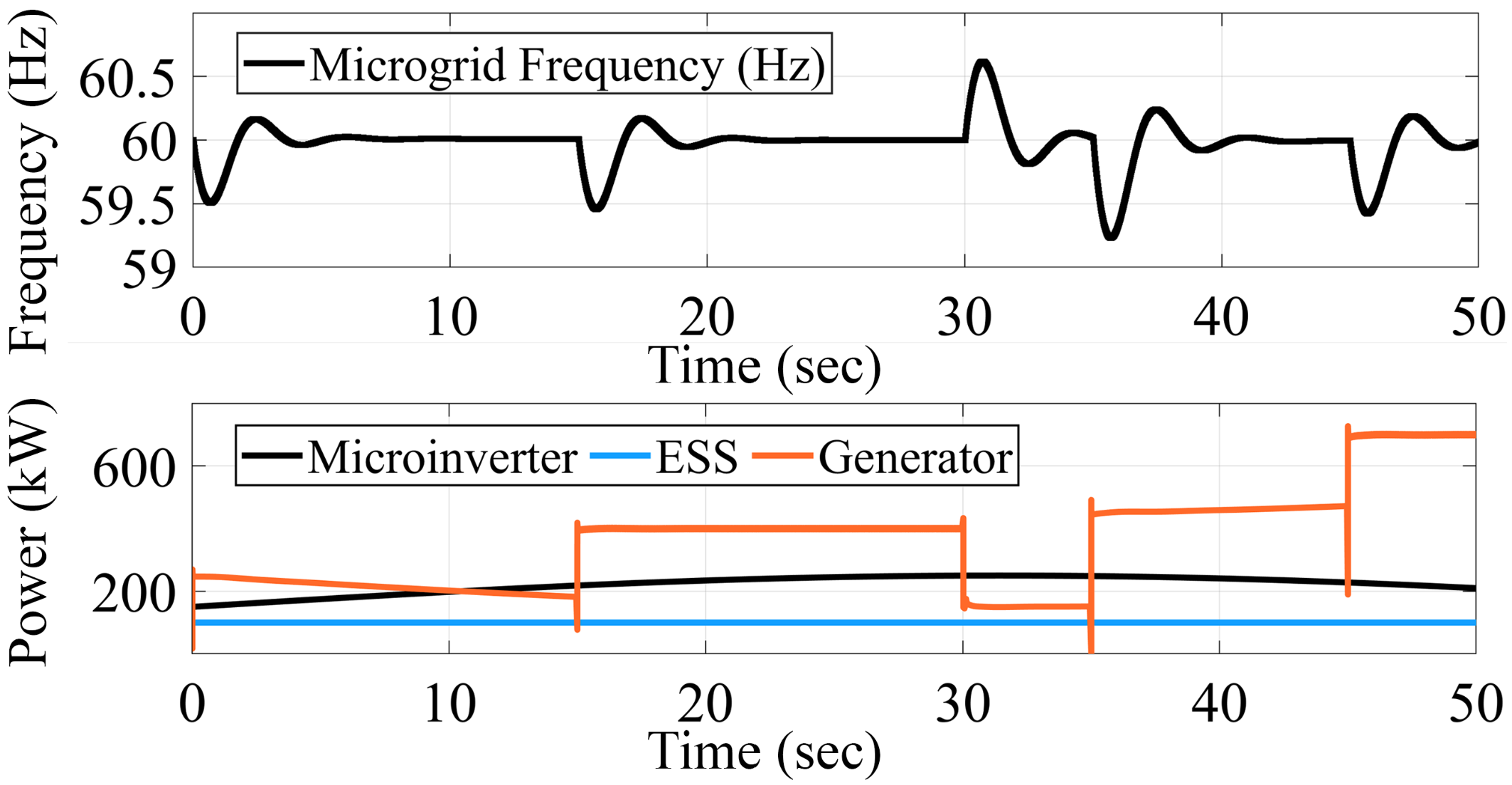}}
\caption{MG behavior under microinverter DoS attack.}
\label{fig:Dos}
\end{figure}

\subsubsection{DoS Attack on the Microinverter}
\label{DOSonMI}
In this case study, we demonstrate a DoS attack in which we disable the output (grid-tied) stage of the microinverter. The outcome of the attack is presented in Figure~\ref{fig:Dos}. As a result of the DoS on the solar microinverter and the unavailability of the solar PV to produce power, the ancillary generation sources are required to meet power demands. In our experiment, a DoS attack disables the microinverter at $t=15~sec$. The inverter  gets back online at $t=30~sec$ immediately before the load increases (at $t=35~sec$), as per the base case scenario (Figure~\ref{fig:nominal}). The microinverter remains connected to the MG until $t=45~sec$ when a DoS attack switches it off again. In this case study, the MG operation is severely impacted due to the frequency fluctuations. Specifically, the malicious control of the inverter in conjunction with the generator switching, which is operated complementarily to meet the MG's power demand, cause frequency instabilities and adversely affect the grid's power quality~\cite{gridImpact}. The impact could vary from damaged equipment, load shedding, brownouts, and even total loss of power (blackout) in case this attack occurs during a peak-load period when the alternative generation sources cannot balance the power demand. This is evident by the widespread power outage across the U.K. in 2019~\cite{UK}.

\subsubsection{Input Tampering Attack on the MPPT Controller}
\label{AttackonMPPTInput}

\begin{figure}[t]
\centerline{\includegraphics[width=0.9\linewidth]{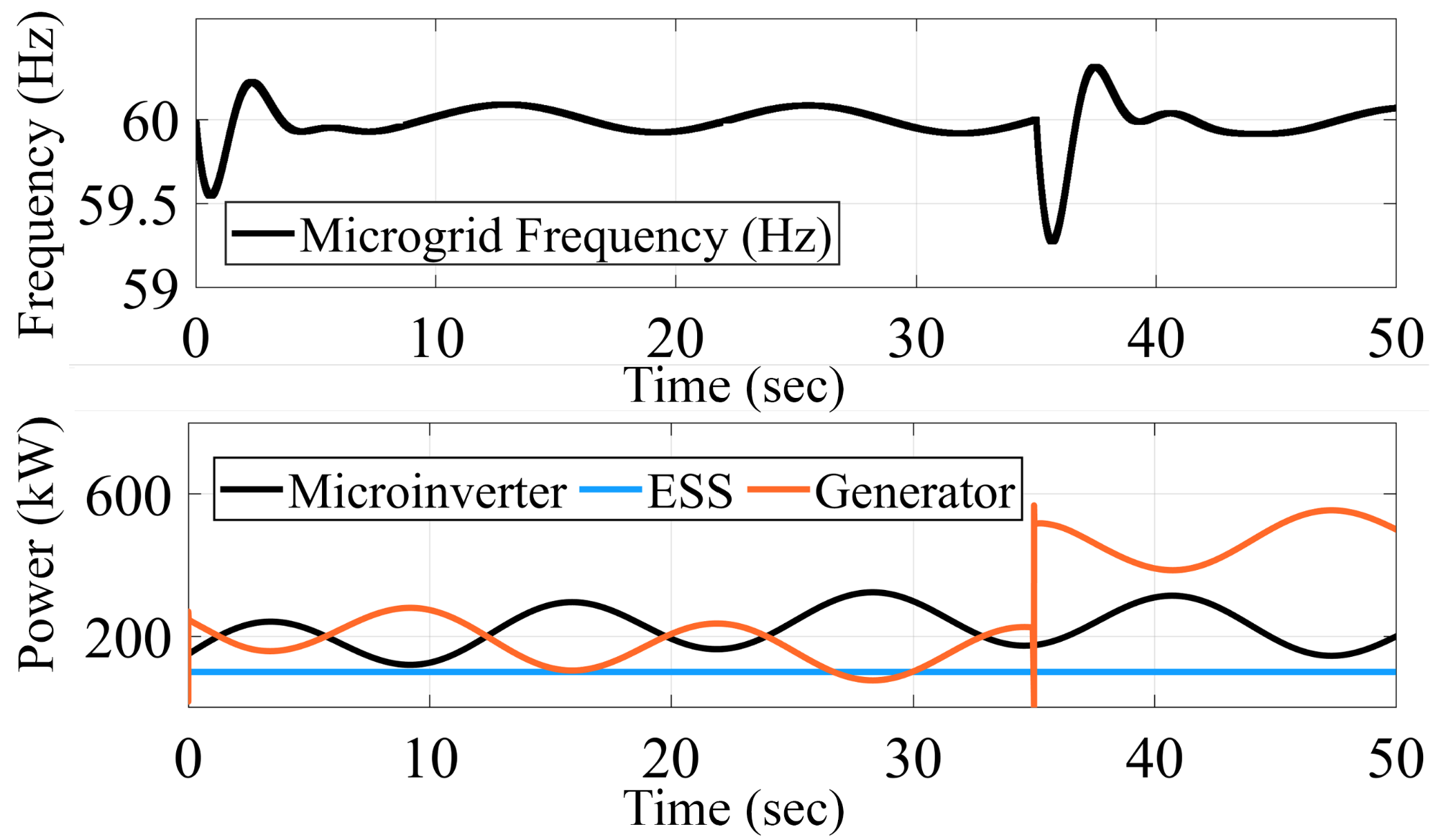}}
\caption{MG behavior under MPPT reference input perturbations.} 
\label{fig:SineAttack}
\end{figure}

\begin{figure}[t]
\centerline{ \includegraphics[width=0.9\linewidth]{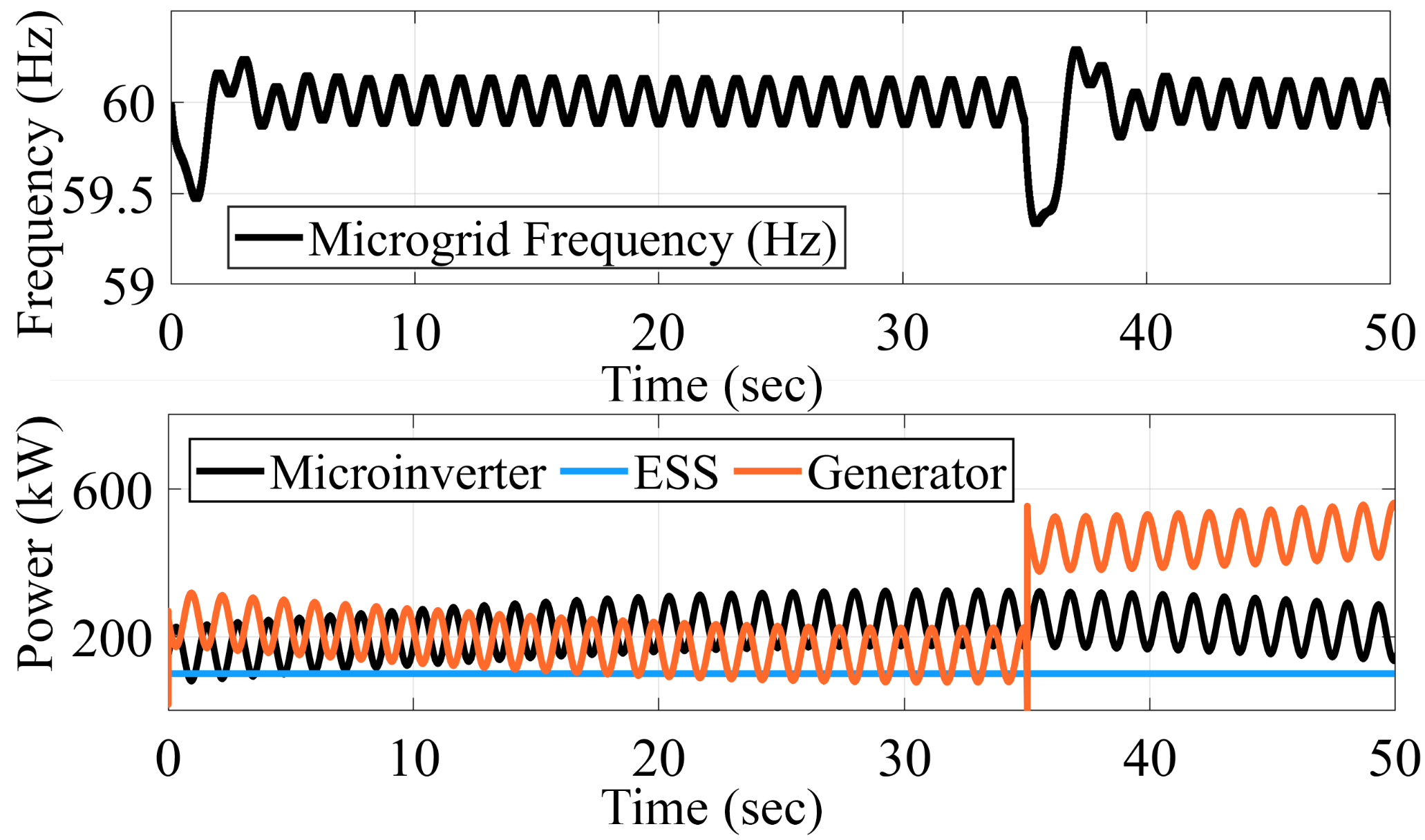}}
\caption{MG behavior under \emph{high frequency} MPPT reference input perturbations.} \label{fig:fastSineAttack}
\end{figure}

As discussed in Section~\ref{s:MPPTcontroller}, the real-time voltage and current measurements of the solar PV module are critical for the efficient operation of the microinverter. The measurements are used by the MPPT in an effort to reach the optimal condition and extract as much power as possible from the PVs. 
Thus, by strategically and stealthily tampering the $V_{rt}$ and $I_{rt}$ sensed values, attackers can destabilize the MPPT controller and consequently the microinverter power generation~\cite{gridImpact}. The severity of this attack scenario is illustrated in Figure~\ref{fig:SineAttack} and Figure~\ref{fig:fastSineAttack}, where two error signals with different frequencies are superimposed over the sensor measurements. In both cases, the applied signal perturbations are constantly moving the operation point (around the optimal one) without allowing the MPPT algorithm to converge and hence, creating an unregulated output behavior. 


By compromising the sensed measurements, the effect on the MG power generation is twofold: first, the microinverter's output becomes unstable, and second, the power deficit needs to be met by alternative power generation units regulating the unstable inverter output.
The latter can either be economically impractical or infeasible since the inertia of rotating generators cannot meet this abnormal pattern~\cite{inertia}. As a result, frequency fluctuations and power outages might occur. If the maliciously compromised microinverter is not promptly isolated from the MG (defensive islanding), the introduced power quality degradation can jeopardize sensitive grid connected devices (e.g., sensing and data acquisition meters) and loads (e.g., industrial and commercial).

\subsection{Attacks Detection Capability} 
\label{sec:Experiment}
The prompt detection of malware in grid assets' firmware code is of paramount importance in preventing any of the previously discussed adverse scenarios. In this section, we present the efficacy of our hardware-assisted methodology utilizing HPCs in detecting malicious firmware modifications. The detection accuracy of the proposed approach renders it a viable proposition for real-world power grid applications.

\subsubsection{Implementation Setup}
\label{exp:setup}
To demonstrate the effectiveness of our approach, we test our
technique with a real-world commercial firmware of an embedded microinverter system~\cite{TMDSSOLA47:online}. Specifically, the TI solar microinverter supports a {\it F2803x} microcontroller, which we utilized for our testing and evaluation phase. 
We employ our custom-built HPCs in order to detect the firmware modifications within the microinverter's code. 
Four firmware modifications are realized into the solar microinverter code according to the attack scenarios simulated in Section~\ref{MGImpact}, i.e., a DoS targeted attack on the MPPT functionality, a DoS attack on the whole microinverter system, and two MPPT input manipulation attacks. 
The attacks are flashed into the {\it Piccolo TMS320F28035 Isolated ControlCARD } using a {\it DIMM100 Docking Station Baseboard} supporting  power and JTAG capabilities.  
To generate the HPC-based signatures, 
a {\it Python } script is utilized to count the number of custom HPCs in the dissembled firmware executable. The {\it TI dis2000} dissembler was used to get the assembly code from the binary executable. Since HPCs are counted in intervals, our script collects the total value for each proposed HPC every $50$ assembly instructions. We utilized this sampling rate as it is large enough to capture information about the structure of the program but small enough to furnish an adequate amount of samples. We utilized a 70:30 spilt where we trained on 70\% of the dataset with the remaining 30\% used for testing. We aggregate samples from all the attacks and the base code. With the attained simulated custom HPCs, we build three ML classifiers generated using the {\it scikit learn} library~\cite{scikit-learn}, a DT, a NN and a RF, as described in Section~\ref{MLclassifiers}.  


\subsubsection{Detecting Firmware Modifications with ML Models}
\label{results-1}

\begin{figure}[t]
\centering
  \includegraphics[width=\linewidth]{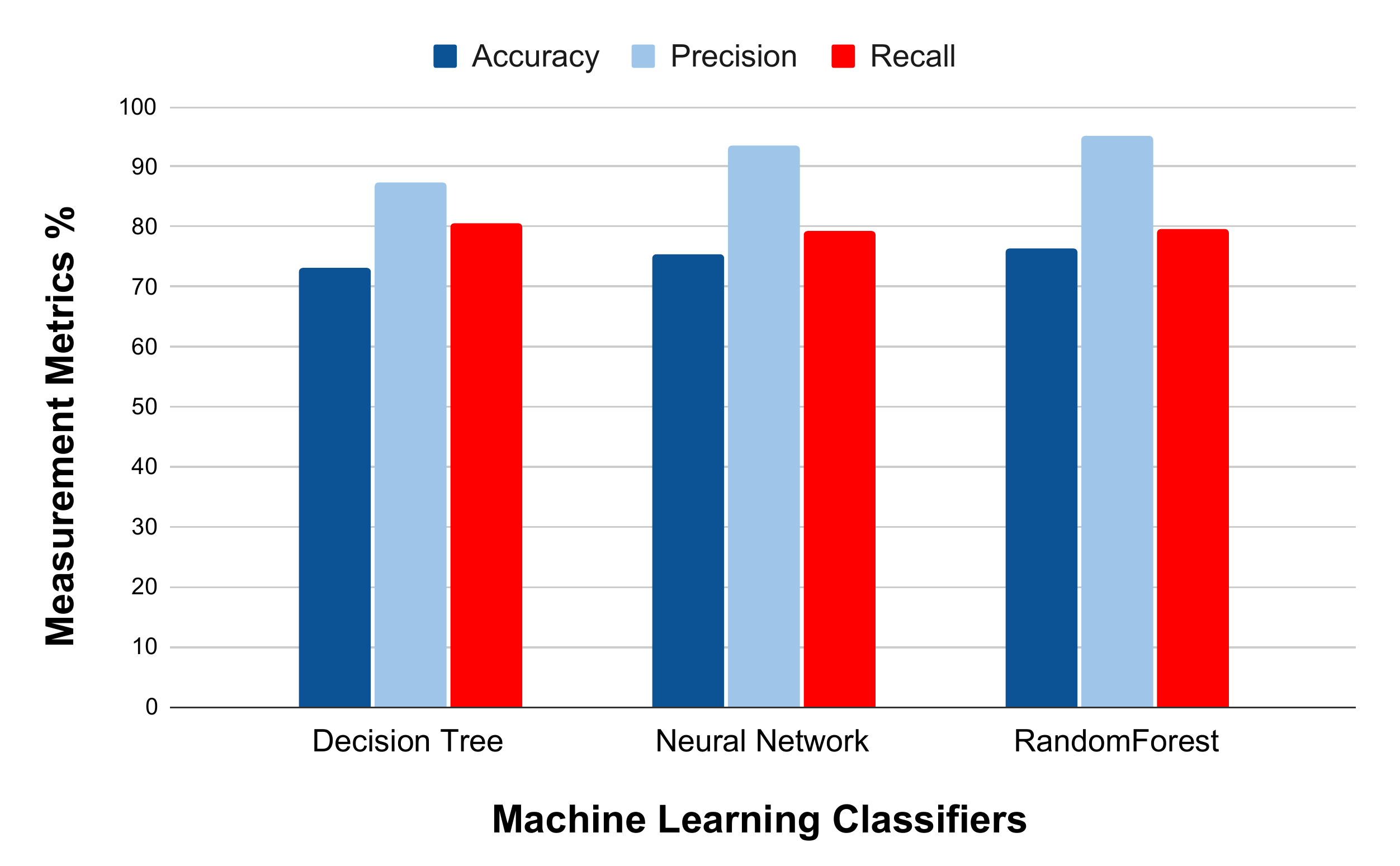}
  \caption{Malicious firmware detection analysis using all $30$ HPCs.}
  \label{fig:HPC1chart}
\end{figure}

In order to detect the simulated attacks on the MG model, we use all the $30$ custom-built HPCs 
to construct the ML classifiers as shown in Figure~\ref{fig:HPC1chart}. It is observed that each model is capable of identifying the firmware modifications with high precision. The best measurement metrics are attained with the RF classifier, which provides an accuracy of 76.4\%, precision of 95.1\%, and recall of 79.53\%. The accuracy and recall for all models is partially diminished because the training dataset has only one benign base version of the solar microinverter code and four malicious firmware variations.

\begin{figure}[t]
\centering
  \includegraphics[width=\linewidth]{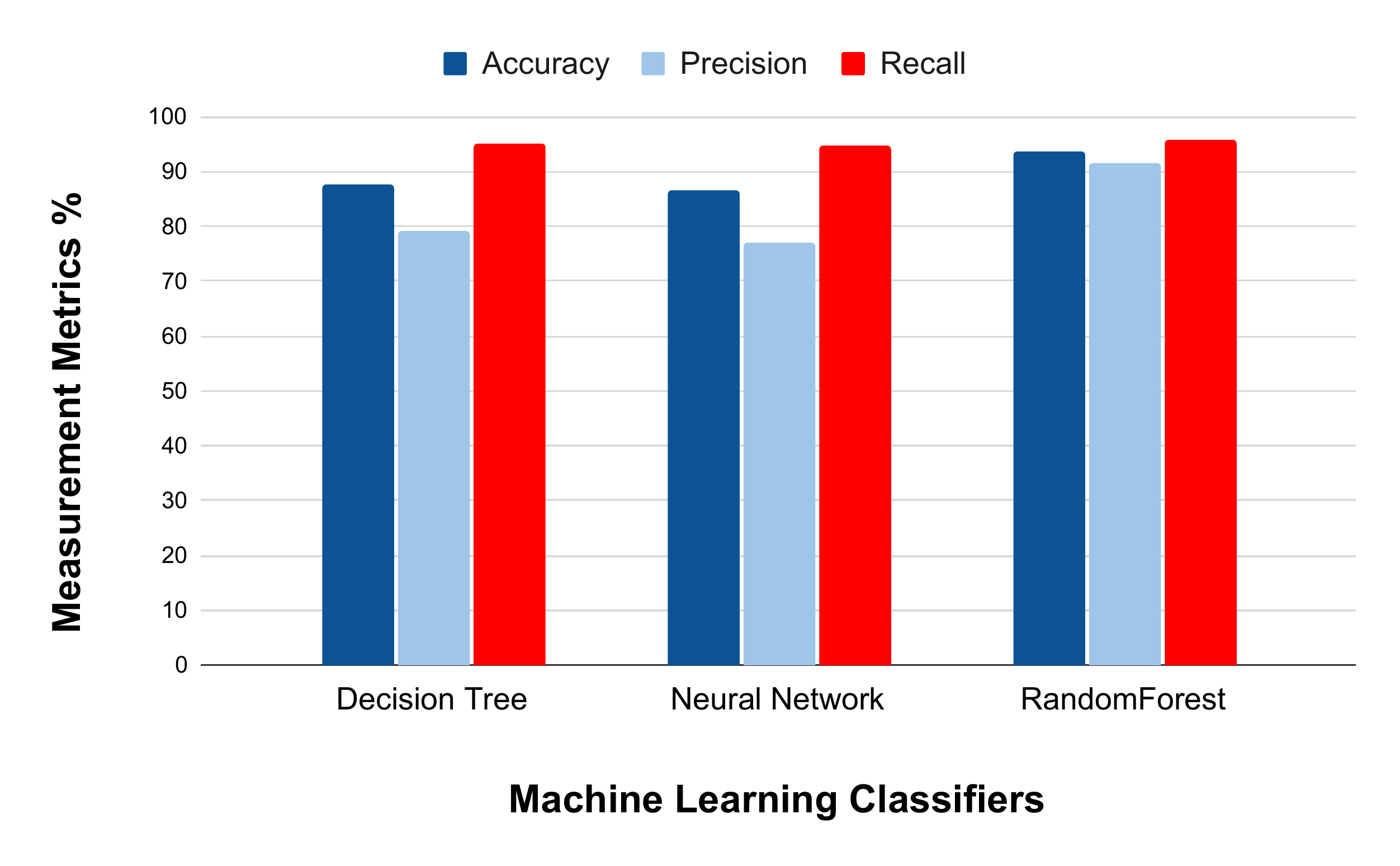}
  \caption{Malicious firmware detection analysis with balanced dataset.}
  \label{fig:revisionchart1}
\end{figure}

{
The imbalance of the dataset can lead to a biased ratio of false positives and false negatives. Consequently, we balance the number of benign and malicious data and retrain our classifiers as shown in Figure~\ref{fig:revisionchart1} with the Random Forest furnishing 93.75\% accuracy, 91.66\% precision, and 95.65\% recall. 
Among the misclassified samples, the false positive to false negative percent ratios are: 15.7\% to 2.5\% for DT, 19.6\% to 7.5\% for NN, and 8\% to 4.5\% for RF. The low number of false negatives indicates the robustness of our proposed methodology in identifying the malicious inverter attacks.} 

{
When analyzing the effects of false positives adversely affecting performance, we consider two particular points. First, from Figure~\ref{fig:revisionchart1}, our proposed methodology is successfully able to identify 93.75\% of samples. Therefore, the number of false positives is relatively low, thereby leading to infrequent disruptions in inverter performance. Second, when ensuring system security for inverters, the minuscule amount of false positives is an acceptable trade-off for identifying nearly all malicious samples. As a result, when considering the strengthened security obtained and the infrequent negative effects on performance, our proposed methodology provides a robust defense. 
}


\begin{figure}[t]
\centering
  \includegraphics[width=\linewidth]{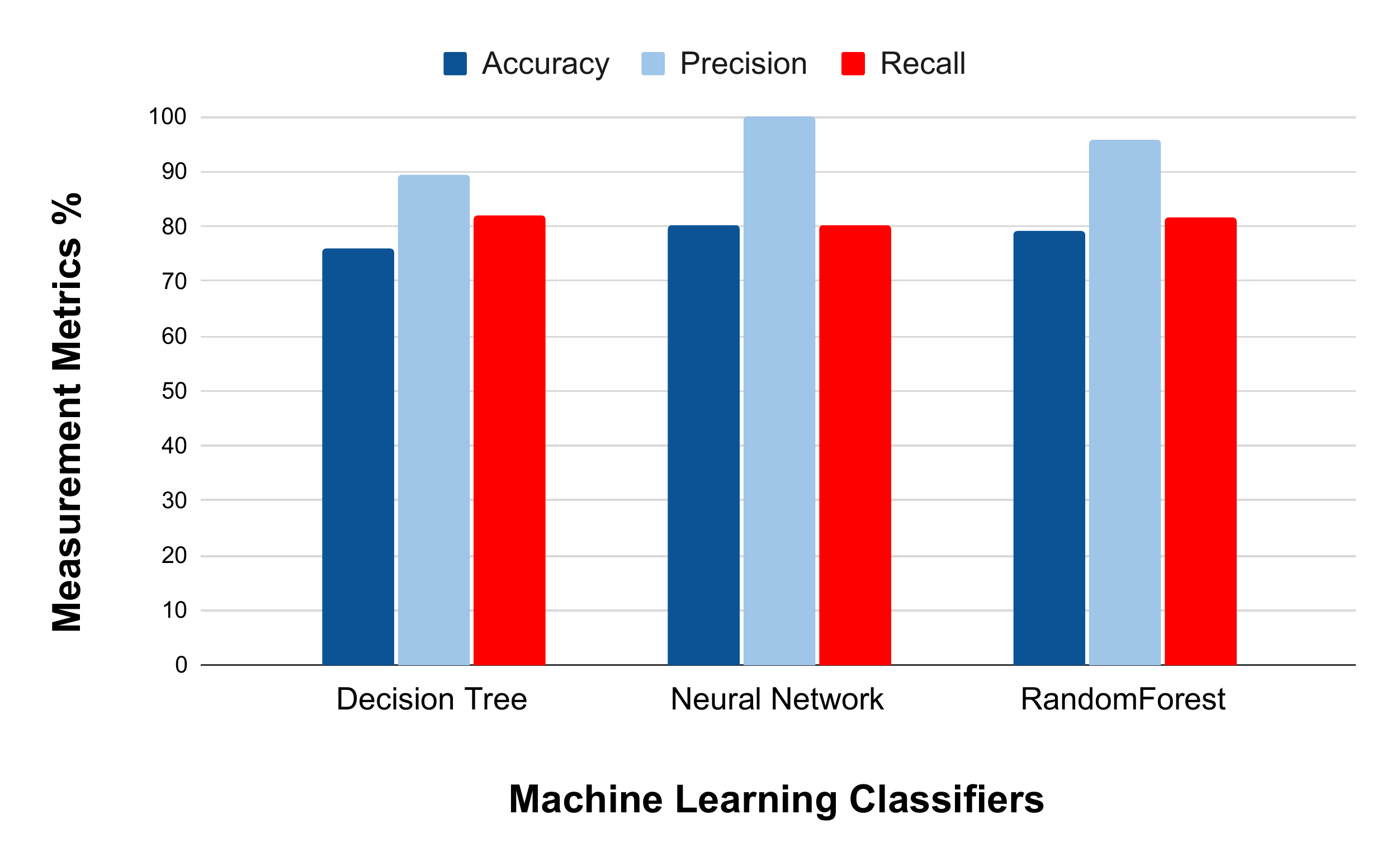}
  \caption{Malicious firmware detection analysis using only three HPCs.}
  \label{fig:HPC2chart}
\end{figure}

\subsubsection{PCA Feature Selection in ML Models}
\label{results-2}

Since most modern processors are only capable of tracking a couple of HPCs at a time, we used PCA,    
as described in Section~\ref{PCA-explanation},
to determine the top best features: boolean, arithmetic and branch instructions ($n$, $a$, and $b$). {These features are robust for malicious firmware detection, as they can adequately capture many of the malignant tendencies found in pernicious firmware. Compromised firmware will attempt to deviate from the normal control flow of the system which increments branch instructions and engenders malicious activity such as manipulating outputs or internal system functions which increments boolean and arithmetic instructions.} Moreover, reducing the number of custom-built HPCs aid in decreasing the hardware overhead of the DfS architecture. These three dominant features are used to train a new set of models as shown in Figure~\ref{fig:HPC2chart}. Experiments show that the performance metrics for each classifier are bolstered by the application of PCA. The DT and RF classifiers have a peak accuracy of 75.84\% and 79.21\%, respectively, and peak precision of  89.51\% and 95.8\%, respectively.
The NN classifier furnishes the best results since the accuracy improves from 75.28\% to 80.33\% and the  precision improves from  93.7\% to 100\%. 
While all models could detect the firmware attacks, the NN is able to detect, with high accuracy and precision, all the malicious firmware samples. This demonstrates the benefit of PCA feature selection utilized in our models. {When utilizing $30$ HPCs, the model becomes overfitted, but employing PCA enabled us to attain the optimal set of HPCs required to construct an effective model.} Only three custom HPCs are sufficient to provide a robust defense against malicious firmware modifications, thus, significantly reducing the DfS hardware overhead.

\begin{table*}[h!t]
\caption{Overhead of individual classifiers.}
\centering
\vspace{0.1in}
\begin{tabular}{||l|l|l|l||}
\hline \hline
\multicolumn{2}{||c|}{\textbf{Design}}                             & \multicolumn{1}{|c|}{{\textbf{\textit{lsi\_10k Library}}}} & \multicolumn{1}{|c||}{\textbf{\textit{saed\_90nm Library}}}    \\ \hline \hline 
\multirow{2}{*}{{Decision Tree}}  & {Area Overhead} & 2108 sq. units    & 8930.814878 sq. units   \\ \cline{2-4} 
 & {Power Overhead} & 44.8547 $\mu$W  & 478.9462 $\mu$W \\ \hline
\multirow{2}{*}{{Neural Network}} & {Area Overhead} & 37062 sq. units   & 153799.401513 sq. units \\ \cline{2-4} 
 & {Power Overhead} & 343.7231 $\mu$W & 5429.1 $\mu$W   \\ \hline
\multirow{2}{*}{{Random Forest}}  & {Area Overhead} & 2618 sq. units    & 11004.428258 sq. units  \\ \cline{2-4} 
 & {Power Overhead} & 47.7563 $\mu$W  & 529.3894 $\mu$W \\ \hline \hline
\end{tabular}
\label{tab:my-table1}
\vspace{-2mm}
\end{table*}

\subsubsection{Hardware Implementation}\label{results-hardware}

{
We realize the ML classifiers of Figure~\ref{fig:HPC2chart} in hardware to determine their area and power overhead. We design the register transfer level (RTL) for the DT, NN, and RF classifiers and synthesize them through the Synopsys Design Compiler logic synthesis tool. We  employ two digital standard cell libraries, \textit{saed\_90nm} and \textit{lsi\_10k}. While other libraries could have been employed, the furnished power and area overhead numbers would be consistent with the values obtainable from our utilized libraries. Our results related to the power and area overhead acquired from our synthesis of the RTL
are presented in Table~\ref{tab:my-table1}. The obtained results are reasonable expenditures for low-end embedded systems, as the power required is in the $\mu$W range. Furthermore, we implement our proposed custom-built HPCs into a simple MIPS processor using Xilinx Vivado high-level synthesis (HLS), which enables C to RTL conversion, and synthesized it for an Artix-7 FPGA. The MIPS processor is chosen as it is open  source and incurs less area overhead compared to Intel or ARM processors. Hence, the area overhead for our custom-built HPC design units would be lower in an Intel or ARM processor than in MIPS. Our custom-built HPC logic is added to the decode unit of the MIPS processor, and we incurred a maximum area overhead of 1.2\% in terms of LUTs. Consequently, implementing our custom-built HPCs into an ARM or Intel processor would result in less overhead.}

{Since inverters are part of the critical infrastructure, the security objective of their underlined embedded systems is of utmost importance. Even our worst-case scenario power consumption overheads range from a few $\mu$W to tenths of mWs. Microinverter controllers are not considered low-power nor battery-operated embedded systems, and such small power trade-offs are acceptable to enhance device security~\cite{10.1145/996566.996771}. The proposed overheads can be compared against similar  microcontroller unit designs. The open core OpenMSP430 design layout, discussed in \cite{PauloJoaoTemoteoRito_6_2018}, considers an embedded system occupying an area of 1050x2800$\mu$m while consuming 1.99mW of power. On the other hand, the authors in~\cite{yang2020pcb} present a DC-AC inverter module occupying a vast printed circuit board (PCB) area of 90×40mm. The reported design specifications render our 1.2\% overhead totally insignificant, and a reasonable sacrifice for a more secure device.}

\subsubsection{Instruction Elimination Analysis}\label{results-3}

\begin{figure}[!t]
\centering
  \includegraphics[width=1\linewidth]{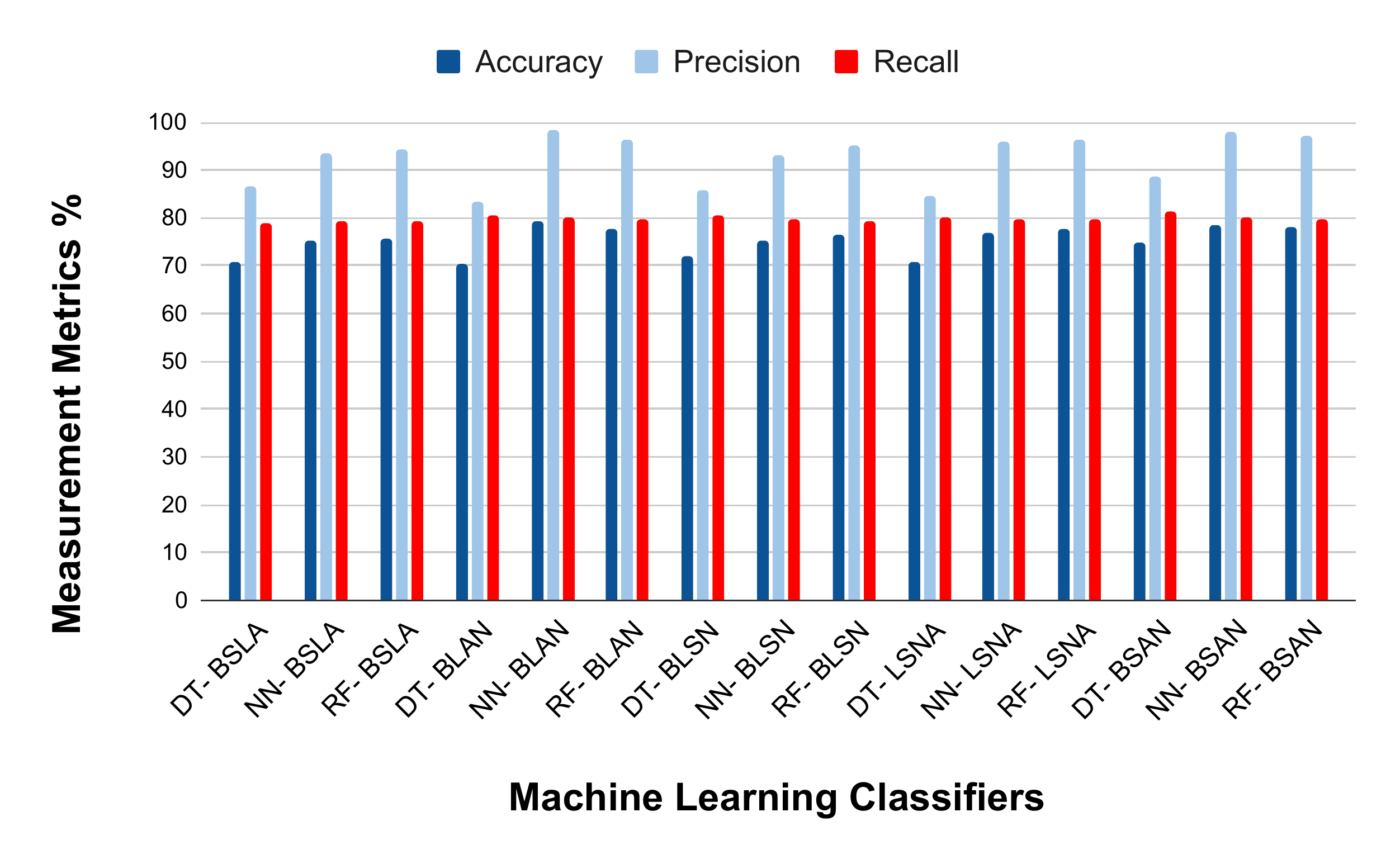}
  \vspace{-1mm}
  \caption{Performance with one excluded instruction.}
  \vspace{-1mm}
  \label{fig:HPC3chart}
\end{figure}

In this experiment, we train ML models on datasets in which any one of the five instructions, i.e., $a$, $s$, $l$, $n$, and $b$,  are excluded at a time. As an example, $BLAN$ represents the scenario in which only branch, load, arithmetic and boolean instructions are included.  As a result, HPCs like $sa$ and $sb$ are not considered in $BLAN$. Figure~\ref{fig:HPC3chart} shows the performance comparison for the ML classifiers on all possible omission combinations. The best results are furnished by the NN classifier trained on the $BLAN$ dataset. The performance metrics of 79.2\% accuracy and 98.6\% precision are the highest throughout all the models trained. This is significant because the experiments presented in Section~\ref{results-2} have shown that boolean, arithmetic and branch instructions are by far the most dominant.

\begin{figure}[t]
\centering
  \includegraphics[width=1\linewidth]{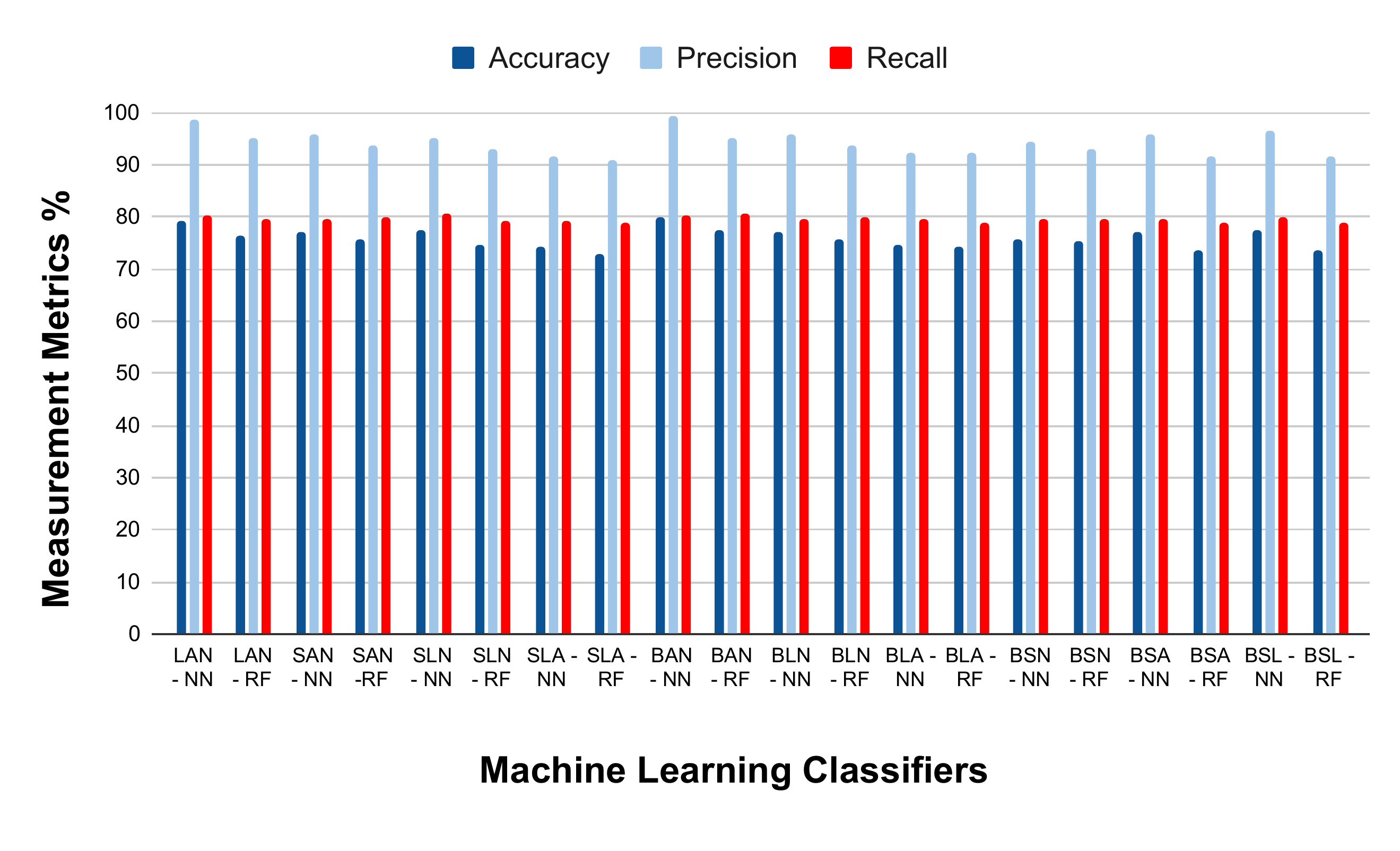}
  \vspace{-1mm}
  \caption{Performance with two excluded instructions.}
  \vspace{-1mm}
  \label{fig:HPC4chart}
\end{figure}

In order to further prove the strength of $n$, $a$, and $b$ instructions, we  trained a final set of models on datasets in which two of the five instructions are excluded. That is, $BAN$ indicates a dataset in which store and load instructions are omitted. Figure~\ref{fig:HPC4chart} shows the classification performance attained from all possible instruction datasets. The NN trained on the $BAN$ dataset has the highest precision of 99.3\%, which further supports our previous results. 
Therefore, HPC support for only boolean, arithmetic, and branch 
instructions can assist in providing better DfS for embedded microinverter controllers.  




\vspace{-1mm}
\section{Conclusion} \label{sec:Conclusion}

In this paper, we demonstrate the impact of firmware modification attacks on a solar microinverter operating in a MG setup. We propose the detection of such stealthy attacks by leveraging custom-built HPCs as DfS and periodically sampling the instructions within the inverter's firmware. Our proof-of-concept validates that devices without native HPC support can be hardened against adversaries leveraging DfS primitives. Specifically, we improve the security of TI's F2803x microcontroller leveraging custom-built HPCs. Our experiments demonstrate that ML models trained on our HPCs identify firmware attacks with high accuracy and precision. Apart from providing an approach to secure solar microinverters without HPC support, we demonstrate the potential impact of compromised devices in MG deployments operating autonomously.

In the future, we will include custom-built HPC extensions for other firmware-controlled grid assets and formalize our detection methodology for additional firmware attacks (e.g., rootkits, command injection, etc.). We plan to evaluate our framework utilizing hardware-in-the-loop experiments including grid-tied inverters and MG components, refine and improve educational and laboratory tutorials \cite{9130868}, and demonstrate the impact of these attacks in real-time operation. 

\vspace{-1mm}

\bibliography{references.bib}

\begin{thebibliography}{10}
\providecommand{\url}[1]{#1}
\csname url@samestyle\endcsname
\providecommand{\newblock}{\relax}
\providecommand{\bibinfo}[2]{#2}
\providecommand{\BIBentrySTDinterwordspacing}{\spaceskip=0pt\relax}
\providecommand{\BIBentryALTinterwordstretchfactor}{4}
\providecommand{\BIBentryALTinterwordspacing}{\spaceskip=\fontdimen2\font plus
\BIBentryALTinterwordstretchfactor\fontdimen3\font minus
  \fontdimen4\font\relax}
\providecommand{\BIBforeignlanguage}[2]{{%
\expandafter\ifx\csname l@#1\endcsname\relax
\typeout{** WARNING: IEEEtran.bst: No hyphenation pattern has been}%
\typeout{** loaded for the language `#1'. Using the pattern for}%
\typeout{** the default language instead.}%
\else
\language=\csname l@#1\endcsname
\fi
#2}}
\providecommand{\BIBdecl}{\relax}
\BIBdecl

\bibitem{Cal_RPS}
\BIBentryALTinterwordspacing
{California Public utilities Commission}, ``{Renewables Portfolio Standard
  (RPS) Program},'' may 2020. [Online]. Available:
  \url{https://www.cpuc.ca.gov/renewables/}
\BIBentrySTDinterwordspacing

\bibitem{iea}
\BIBentryALTinterwordspacing
I.~E. Agency, ``Market analysis and forecast from 2019 to 2024,'' May 2020.
  [Online]. Available:
  \url{https://www.iea.org/reports/renewables-2019/distributed-solar-pv#abstract}
\BIBentrySTDinterwordspacing

\bibitem{electrify}
\BIBentryALTinterwordspacing
{Huang, Z. and Krishnaswami, H. and Yuan G. and Huang, R.}, ``{IEEE
  Electrification Magazine: Ubiquitous Power Electronics in Future Power
  Systems},'' September 2020. [Online]. Available: \url{https://bit.ly/2FuM7si}
\BIBentrySTDinterwordspacing

\bibitem{ZHOU201630}
\BIBentryALTinterwordspacing
B.~Zhou, W.~Li, K.~Wing~Chan, Y.~Cao, Y.~Kuang, X.~Liu, and X.~Wang, ``{Smart
  home energy management systems: Concept, configurations, and scheduling
  strategies},'' \emph{Renewable and Sustainable Energy Reviews}, vol.~61, pp.
  30 -- 40, 2016. [Online]. Available:
  \url{http://www.sciencedirect.com/science/article/pii/S1364032116002823}
\BIBentrySTDinterwordspacing

\bibitem{han2014smart}
J.~Han, C.-S. Choi, W.-K. Park, I.~Lee, and S.-H. Kim, ``{Smart home energy
  management system including renewable energy based on ZigBee and PLC},''
  \emph{IEEE Transactions on Consumer Electronics}, vol.~60, no.~2, pp.
  198--202, 2014.

\bibitem{IEEE1547}
I.~1547, ``{IEEE Standard Conformance Test Procedures for Equipment
  Interconnecting Distributed Energy Resources with Electric Power Systems and
  Associated Interfaces},''
  \url{https://standards.ieee.org/standard/1547_1-2020.html}, 2020, accessed:
  2020-05-27.

\bibitem{sayghe2020survey}
A.~Sayghe, Y.~Hu, I.~Zografopoulos, X.~Liu, R.~G. Dutta, Y.~Jin, and
  C.~Konstantinou, ``Survey of machine learning methods for detecting false
  data injection attacks in power systems,'' \emph{IET Smart Grid}, vol.~3, pp.
  581--595(14), October 2020.

\bibitem{roadmap}
J.~Johnson, ``Roadmap for photovoltaic cyber security,'' 12 2017.

\bibitem{zografopoulos2020derauth}
I.~Zografopoulos and C.~Konstantinou, ``{DERauth: A Battery-based
  Authentication Scheme for Distributed Energy Resources},'' in \emph{2020 IEEE
  Computer Society Annual Symposium on VLSI (ISVLSI)}.\hskip 1em plus 0.5em
  minus 0.4em\relax IEEE, 2020, pp. 560--567.

\bibitem{IEEE1815}
1815.1-2015, ``{IEEE Standard for Exchanging Information Between Networks
  Implementing IEC 61850 and IEEE Std 1815(TM) [Distributed Network Protocol
  (DNP3)]},'' 2020.

\bibitem{zografopoulos2020harness}
I.~Zografopoulos, J.~Ospina, and C.~Konstantinou, ``Special session: Harness
  the power of ders for secure communications in electric energy systems,'' in
  \emph{2020 IEEE 38th International Conference on Computer Design
  (ICCD)}.\hskip 1em plus 0.5em minus 0.4em\relax IEEE, 2020, pp. 49--52.

\bibitem{remoteDanger}
P.~Fairley, ``{800,000 Microinverters Remotely Retrofitted on Oahu—in One
  Day},''
  \url{https://spectrum.ieee.org/energywise/green-tech/solar/in-one-day-800000-microinverters-remotely-retrofitted-on-oahu},
  2015, accessed: 2020-05-15.

\bibitem{stright2020defensive}
J.~Stright, P.~Cheetham, and C.~Konstantinou, ``Defensive cost-benefit analysis
  of smart grid digital functionalities,'' \emph{arXiv preprint
  arXiv:2008.12843}, 2020.

\bibitem{9107609}
H.~{Benkraouda}, M.~A. {Chakkantakath}, A.~{Keliris}, and M.~{Maniatakos},
  ``Snifu: Secure network interception for firmware updates in legacy plcs,''
  in \emph{2020 IEEE 38th VLSI Test Symposium (VTS)}, 2020, pp. 1--6.

\bibitem{falas2020modular}
S.~Falas, C.~Konstantinou, and M.~K. Michael, ``A modular end-to-end framework
  for secure firmware updates on embedded systems,'' \emph{arXiv preprint
  arXiv:2007.09071}, 2020.

\bibitem{qi2016cybersecurity}
J.~Qi, A.~Hahn, X.~Lu, J.~Wang, and C.~Liu, ``Cybersecurity for distributed
  energy resources and smart inverters,'' \emph{IET Cyber-Physical Systems:
  Theory \& Applications}, vol.~1, no.~1, pp. 28--39, 2016.

\bibitem{8975537}
T.~S. {Ustun}, ``Cybersecurity vulnerabilities of smart inverters and their
  impacts on power system operation,'' in \emph{2019 International Conference
  on Power Electronics, Control and Automation (ICPECA)}, 2019, pp. 1--4.

\bibitem{Trisis1}
D.~Inc., ``{TRISIS Malware-Analysis of Safety System Targeted Malware},''
  \url{https://www.dragos.com/wp-content/uploads/TRISIS-01.pdf}, 2018,
  accessed: 2020-05-15.

\bibitem{murad2010evading}
K.~Murad, S.~Shirazi, Y.~Zikria, and N.~Ikram, ``Evading virus detection using
  code obfuscation,'' in \emph{International Conference on Future Generation
  Information Technology}, Springer.\hskip 1em plus 0.5em minus 0.4em\relax
  Heidelberg, Germany: Springer, Berlin, Heidelberg, 2010, pp. 394--401.

\bibitem{demme2013feasibility}
J.~Demme, M.~Maycock, J.~Schmitz, A.~Tang, A.~Waksman, S.~Sethumadhavan, and
  S.~Stolfo, ``On the feasibility of online malware detection with performance
  counters,'' \emph{ACM SIGARCH Computer Architecture News}, vol.~41, no.~3,
  pp. 559--570, 2013.

\bibitem{malone2011hardware}
C.~Malone, M.~Zahran, and R.~Karri, ``Are hardware performance counters a cost
  effective way for integrity checking of programs,'' in \emph{Proceedings of
  the sixth ACM workshop on Scalable trusted computing}.\hskip 1em plus 0.5em
  minus 0.4em\relax Chicago Illinois USA: ACM, 2011, pp. 71--76.

\bibitem{ozsoy2016hardware}
M.~Ozsoy, K.~N. Khasawneh, C.~Donovick, I.~Gorelik, N.~Abu-Ghazaleh, and
  D.~Ponomarev, ``Hardware-based malware detection using low-level
  architectural features,'' \emph{IEEE Transactions on Computers}, vol.~65,
  no.~11, pp. 3332--3344, 2016.

\bibitem{sayadi20192smart}
H.~Sayadi, H.~M. Makrani, S.~M.~P. Dinakarrao, T.~Mohsenin, A.~Sasan,
  S.~Rafatirad, and H.~Homayoun, ``2smart: A two-stage machine learning-based
  approach for run-time specialized hardware-assisted malware detection,'' in
  \emph{2019 Design, Automation \& Test in Europe Conference \& Exhibition
  (DATE)}, IEEE.\hskip 1em plus 0.5em minus 0.4em\relax Florence, Italy, Italy:
  IEEE, 2019, pp. 728--733.

\bibitem{wang2016malicious}
X.~Wang, C.~Konstantinou, M.~Maniatakos, R.~Karri, S.~Lee, P.~Robison,
  P.~Stergiou, and S.~Kim, ``Malicious firmware detection with hardware
  performance counters,'' \emph{IEEE Transactions on Multi-Scale Computing
  Systems}, vol.~2, no.~3, pp. 160--173, 2016.

\bibitem{NISTsmartgrid}
\BIBentryALTinterwordspacing
NIST, ``{NIST \& The Smart Grid},'' May 2020. [Online]. Available:
  \url{https://www.nist.gov/el/smart-grid/about-smart-grid/nist-and-smart-grid}
\BIBentrySTDinterwordspacing

\bibitem{smartgrid}
\BIBentryALTinterwordspacing
Smartgrid.gov, ``The smart grid,'' May 2020. [Online]. Available:
  \url{https://www.smartgrid.gov/the_smart_grid/smart_grid.html}
\BIBentrySTDinterwordspacing

\bibitem{arbab2019smart}
B.~Arbab-Zavar, E.~Palacios-Garcia, J.~Vasquez, and J.~Guerrero, ``Smart
  inverters for microgrid applications: a review,'' \emph{Energies}, vol.~12,
  no.~5, p. 840, 2019.

\bibitem{Inv1}
{ENPHASE}, ``{Comparing Inverters},''
  \url{https://enphase.com/en-us/products-and-services/microinverters/vs-string-inverter},
  2020, accessed: 2020-05-5.

\bibitem{Inv2}
{MPPsolar}, ``{Inverter Selection guide},''
  \url{https://www.mppsolar.com/v3/inverter-selection-guide-2/}, 2020,
  accessed: 2020-05-05.

\bibitem{Inv3}
{Wind \& Sun}, ``{Inverter Basics and Selecting the Right Model},''
  \url{https://www.solar-electric.com/learning-center/inverter-basics-selection.html/}.

\bibitem{InvStatus}
J.~Berdner, ``{NREL Integrating PV in Distribution Grids: Advanced Inverter
  Status, CA \& HI},''
  \url{https://www.nrel.gov/esif/assets/pdfs/highpenworkshop_berdner.pdf},
  2015, accessed: 2020-05-25.

\bibitem{InvEf}
W.~Cha, Y.~Cho, J.~Kwon, and B.~Kwon, ``Highly efficient microinverter with
  soft-switching step-up converter and single-switch-modulation inverter,''
  \emph{IEEE Transactions on Industrial Electronics}, vol.~62, no.~6, pp.
  3516--3523, 2014.

\bibitem{smartDanger}
J.~Sattler, ``{Hypponen’s Law: If it’s smart, it’s vulnerable},''
  \url{https://blog.f-secure.com/hypponens-law-smart-vulnerable/}, 2018,
  accessed: 2020-05-15.

\bibitem{liu2020deep}
X.~{Liu}, J.~{Ospina}, and C.~{Konstantinou}, ``Deep reinforcement learning for
  cybersecurity assessment of wind integrated power systems,'' \emph{IEEE
  Access}, vol.~8, pp. 208\,378--208\,394, 2020.

\bibitem{ICS_attack}
K.~Hemsley and R.~Fisher, ``A history of cyber incidents and threats involving
  industrial control systems,'' in \emph{Critical Infrastructure Protection
  XII}, J.~Staggs and S.~Shenoi, Eds.\hskip 1em plus 0.5em minus 0.4em\relax
  Cham: Springer International Publishing, 2018, pp. 215--242.

\bibitem{mclaughlin2016cybersecurity}
S.~McLaughlin, C.~Konstantinou, X.~Wang, L.~Davi, A.~Sadeghi, M.~Maniatakos,
  and R.~Karri, ``The cybersecurity landscape in industrial control systems,''
  \emph{Proceedings of the IEEE}, vol. 104, no.~5, pp. 1039--1057, 2016.

\bibitem{asiri_2018}
\BIBentryALTinterwordspacing
A.~Sidath, ``Machine learning classifiers,'' Jun 2018. [Online]. Available:
  \url{https://towardsdatascience.com/machine-learning-classifiers-a5cc4e1b0623}
\BIBentrySTDinterwordspacing

\bibitem{HowtoCon88:online}
J.~Browniee, ``How to configure the number of layers and nodes in a neural
  network,'' \url{https://tinyurl.com/yxw47q3s}, May 2019.

\bibitem{Understa84:online}
T.~Yiu, ``Understanding random forest - towards data science,''
  \url{https://towardsdatascience.com/understanding-random-forest-58381e0602d2},
  June 2019.

\bibitem{MachineL58:online}
J.~Hiu, ``Machine learning — singular value decomposition (svd) \& principal
  component analysis (pca),'' \url{https://tinyurl.com/y4cbk3eh}, March 2019.

\bibitem{Attack1}
A.~{Patel} and S.~{Purwar}, ``Destabilizing smart grid by dynamic load altering
  attack using pi controller,'' in \emph{2017 International Conference on
  Intelligent Computing, Instrumentation and Control Technologies (ICICICT)},
  2017, pp. 354--359.

\bibitem{stuxnet}
R.~Langner, ``Stuxnet: Dissecting a cyberwarfare weapon,'' \emph{IEEE Security
  \& Privacy}, vol.~9, no.~3, pp. 49--51, 2011.

\bibitem{zografopoulos2021cyberphysical}
I.~Zografopoulos, J.~Ospina, X.~Liu, and C.~Konstantinou, ``{Cyber-Physical
  Energy Systems Security: Threat Modeling, Risk Assessment, Resources,
  Metrics, and Case Studies},'' \emph{arXiv preprint arXiv:2101.10198}, 2021.

\bibitem{DER2}
C.~{Carter}, I.~{Onunkwo}, P.~{Cordeiro}, and J.~{Johnson}, ``Cyber security
  assessment of distributed energy resources,'' in \emph{2017 IEEE 44th
  Photovoltaic Specialist Conference (PVSC)}, 2017, pp. 2135--2140.

\bibitem{DER0}
J.~{Qi}, A.~{Hahn}, X.~{Lu}, J.~{Wang}, and C.~{Liu}, ``Cybersecurity for
  distributed energy resources and smart inverters,'' \emph{IET Cyber-Physical
  Systems: Theory Applications}, vol.~1, no.~1, pp. 28--39, 2016.

\bibitem{DER1}
J.~{Johnson}, J.~{Quiroz}, R.~{Concepcion}, F.~{Wilches-Bernal}, and M.~J.
  {Reno}, ``Power system effects and mitigation recommendations for der
  cyberattacks,'' \emph{IET Cyber-Physical Systems: Theory Applications},
  vol.~4, no.~3, pp. 240--249, 2019.

\bibitem{li2011viper}
Y.~Li, J.~McCune, and A.~Perrig, ``Viper: verifying the integrity of
  peripherals' firmware,'' in \emph{Proceedings of the 18th ACM conference on
  Computer and communications security}, 2011, pp. 3--16.

\bibitem{maskiewicz2014mouse}
J.~Maskiewicz, B.~Ellis, J.~Mouradian, and H.~Shacham, ``Mouse trap: Exploiting
  firmware updates in $\{$USB$\}$ peripherals,'' in \emph{8th $\{$USENIX$\}$
  Workshop on Offensive Technologies ($\{$WOOT$\}$ 14)}, 2014.

\bibitem{lemay2009cumulative}
M.~LeMay and C.~Gunter, ``Cumulative attestation kernels for embedded
  systems,'' in \emph{European Symposium on Research in Computer
  Security}.\hskip 1em plus 0.5em minus 0.4em\relax Springer, 2009, pp.
  655--670.

\bibitem{abad2013chip}
F.~Abad, J.~Van Der~Woude, Y.~Lu, S.~Bak, M.~Caccamo, L.~Sha, R.~Mancuso, and
  S.~Mohan, ``On-chip control flow integrity check for real time embedded
  systems,'' in \emph{2013 IEEE 1st International Conference on Cyber-Physical
  Systems, Networks, and Applications (CPSNA)}.\hskip 1em plus 0.5em minus
  0.4em\relax IEEE, 2013, pp. 26--31.

\bibitem{wang2015reusing}
X.~Wang and R.~Karri, ``Reusing hardware performance counters to detect and
  identify kernel control-flow modifying rootkits,'' \emph{IEEE Transactions on
  Computer-Aided Design of Integrated Circuits and Systems}, vol.~35, no.~3,
  pp. 485--498, 2015.

\bibitem{krishnamurthy2019anomaly}
P.~Krishnamurthy, R.~Karri, and F.~Khorrami, ``Anomaly detection in real-time
  multi-threaded processes using hardware performance counters,'' \emph{IEEE
  Transactions on Information Forensics and Security}, vol.~15, pp. 666--680,
  2019.

\bibitem{bahador2014hpcmalhunter}
M.~Bahador, M.~Abadi, and A.~Tajoddin, ``Hpcmalhunter: Behavioral malware
  detection using hardware performance counters and singular value
  decomposition,'' in \emph{2014 4th International Conference on Computer and
  Knowledge Engineering (ICCKE)}.\hskip 1em plus 0.5em minus 0.4em\relax IEEE,
  2014, pp. 703--708.

\bibitem{wang2015confirm}
X.~Wang, C.~Konstantinou, M.~Maniatakos, and R.~Karri, ``Confirm: Detecting
  firmware modifications in embedded systems using hardware performance
  counters,'' in \emph{2015 IEEE/ACM International Conference on Computer-Aided
  Design (ICCAD)}, IEEE.\hskip 1em plus 0.5em minus 0.4em\relax Austin, TX,
  USA: IEEE, 2015, pp. 544--551.

\bibitem{ozsoy2015malware}
M.~Ozsoy, C.~Donovick, I.~Gorelik, N.~Abu-Ghazaleh, and D.~Ponomarev,
  ``Malware-aware processors: A framework for efficient online malware
  detection,'' in \emph{2015 IEEE 21st International Symposium on High
  Performance Computer Architecture (HPCA)}, IEEE.\hskip 1em plus 0.5em minus
  0.4em\relax Burlingame, CA, USA: IEEE, 2015, pp. 651--661.

\bibitem{rohan2019can}
A.~Rohan, K.~Basu, and R.~Karri, ``Can monitoring system state+ counting custom
  instruction sequences aid malware detection?'' in \emph{2019 IEEE 28th Asian
  Test Symposium (ATS)}, IEEE.\hskip 1em plus 0.5em minus 0.4em\relax Kolkata,
  India: IEEE, 2019, pp. 61--615.

\bibitem{7741452}
C.~{Konstantinou}, A.~{Keliris}, and M.~{Maniatakos}, ``Taxonomy of firmware
  trojans in smart grid devices,'' in \emph{2016 IEEE Power and Energy Society
  General Meeting (PESGM)}, 2016, pp. 1--5.

\bibitem{konstantinou2019hardware}
C.~Konstantinou and M.~Maniatakos, ``Hardware-layer intelligence collection for
  smart grid embedded systems,'' \emph{Journal of Hardware and Systems
  Security}, vol.~3, no.~2, pp. 132--146, 2019.

\bibitem{konstantinou2015impact}
------, ``Impact of firmware modification attacks on power systems field
  devices,'' in \emph{2015 IEEE International Conference on Smart Grid
  Communications (SmartGridComm)}.\hskip 1em plus 0.5em minus 0.4em\relax IEEE,
  2015, pp. 283--288.

\bibitem{10.1145/2994487.2994491}
------, ``A case study on implementing false data injection attacks against
  nonlinear state estimation,'' in \emph{Proceedings of the 2nd ACM Workshop on
  Cyber-Physical Systems Security and Privacy}, ser. CPS-SPC '16.\hskip 1em
  plus 0.5em minus 0.4em\relax New York, NY, USA: Association for Computing
  Machinery, 2016, p. 81–92.

\bibitem{7523254}
F.~{Khorrami}, P.~{Krishnamurthy}, and R.~{Karri}, ``Cybersecurity for control
  systems: A process-aware perspective,'' \emph{IEEE Design Test}, vol.~33,
  no.~5, pp. 75--83, 2016.

\bibitem{microPerformance}
S.~Harb, M.~Kedia, H.~Zhang, and R.~Balog, ``Microinverter and string inverter
  grid-connected photovoltaic system—a comprehensive study,'' in \emph{2013
  IEEE 39th Photovoltaic Specialists Conference (PVSC)}.\hskip 1em plus 0.5em
  minus 0.4em\relax IEEE, 2013, pp. 2885--2890.

\bibitem{MPPTs}
H.~Bounechba, A.~Bouzid, H.~Snani, and A.~Lashab, ``Real time simulation of
  mppt algorithms for pv energy system,'' \emph{International Journal of
  Electrical Power \& Energy Systems}, vol.~83, pp. 67--78, 2016.

\bibitem{PO_1}
D.~{Sera}, L.~{Mathe}, T.~{Kerekes}, S.~V. {Spataru}, and R.~{Teodorescu}, ``On
  the perturb-and-observe and incremental conductance mppt methods for pv
  systems,'' \emph{IEEE Journal of Photovoltaics}, vol.~3, no.~3, pp.
  1070--1078, 2013.

\bibitem{BIANCONI2013346}
E.~Bianconi, J.~Calvente, R.~Giral, E.~Mamarelis, G.~Petrone, C.~A. Ramos-Paja,
  G.~Spagnuolo, and M.~Vitelli, ``{Perturb and observe MPPT algorithm with a
  current controller based on the sliding mode},'' \emph{International Journal
  of Electrical Power \& Energy Systems}, vol.~44, no.~1, pp. 346--356, 2013.

\bibitem{anubi2019enhanced}
O.~M. {Anubi} and C.~{Konstantinou}, ``Enhanced resilient state estimation
  using data-driven auxiliary models,'' \emph{IEEE Transactions on Industrial
  Informatics}, vol.~16, no.~1, pp. 639--647, 2020.

\bibitem{chen2013economic}
Y.~Chen, S.~Lu, Y.~Chang, T.~Lee, and M.~Hu, ``Economic analysis and optimal
  energy management models for microgrid systems: A case study in taiwan,''
  \emph{Applied Energy}, vol. 103, pp. 145--154, 2013.

\bibitem{gridImpact}
T.~Key and K.~Forsten, ``Security, quality, reliability and availability:
  Metrics definition: Progress report,'' \emph{EPRI. USA}, 2005.

\bibitem{UK}
{BBC News}, ``Major power failure affects homes and transport,'' [Online].
  Available: \url{https://www.bbc.com/news/uk-49300025}, BBC News, August 9,
  2019.

\bibitem{inertia}
N.~Soni, S.~Doolla, and M.~Chandorkar, ``Inertia design methods for islanded
  microgrids having static and rotating energy sources,'' \emph{IEEE
  Transactions on Industry Applications}, vol.~52, no.~6, pp. 5165--5174, 2016.

\bibitem{TMDSSOLA47:online}
T.~Instruments, ``Tmdssolaruinvkit solar micro inverter development kit |
  ti.com,'' \url{http://www.ti.com/tool/TMDSSOLARUINVKIT}, Janurary 2020.

\bibitem{scikit-learn}
F.~Pedregosa, G.~Varoquaux, A.~Gramfort, V.~Michel, B.~Thirion, O.~Grisel,
  M.~Blondel, A.~Müller, J.~Nothman, G.~Louppe, P.~Prettenhofer, R.~Weiss,
  V.~Dubourg, J.~Vanderplas, A.~Passos, D.~Cournapeau, M.~Brucher, M.~Perrot,
  and E.~Duchesnay, ``{Scikit-learn: Machine Learning in Python},''
  \emph{Journal of Machine Learning Research}, vol.~12, pp. 2825--2830, 2011.

\bibitem{10.1145/996566.996771}
\BIBentryALTinterwordspacing
P.~Kocher, R.~Lee, G.~McGraw, A.~Raghunathan, and S.~Ravi, ``Security as a new
  dimension in embedded system design,'' in \emph{Proceedings of the 41st
  Annual Design Automation Conference}, ser. DAC '04.\hskip 1em plus 0.5em
  minus 0.4em\relax New York, NY, USA: Association for Computing Machinery,
  2004, p. 753–760. [Online]. Available:
  \url{https://doi.org/10.1145/996566.996771}
\BIBentrySTDinterwordspacing

\bibitem{PauloJoaoTemoteoRito_6_2018}
P.~J.~T. Rito, ``Soc implementation of openmsp430 microcontroller in {UMC}
  130nm,'' Master's Thesis, Univ. Lisboa, Jun. 2018.

\bibitem{yang2020pcb}
Y.~Yang and A.~Emadi, ``Pcb embedded chip-on-chip packaging of a 48 kw sic
  mosfet dc-ac module with double-side cooling design,'' in \emph{2020 IEEE
  Transportation Electrification Conference \& Expo (ITEC)}.\hskip 1em plus
  0.5em minus 0.4em\relax IEEE, 2020, pp. 912--917.

\bibitem{9130868}
C.~{Konstantinou}, ``Cyber-physical systems security education through hands-on
  lab exercises,'' \emph{IEEE Design Test}, vol.~37, no.~6, pp. 47--55, 2020.

\end{thebibliography}

\bibliographystyle{IEEEtran}


\end{document}